\documentclass[a4paper]{article}

\usepackage{amssymb,amsmath,amsthm,verbatim}
\usepackage[final]{graphicx}
\usepackage{natbib}
\usepackage{hyperref}
\usepackage[normalem]{ulem} 

\usepackage{color}

\newtheorem{theorem}{Theorem}[section]
\newtheorem{lemma}[theorem]{Lemma}
\newtheorem{cor}[theorem]{Corollary}
\newtheorem{prop}[theorem]{Proposition}
\theoremstyle{definition}
\newtheorem{example}[theorem]{Example}
\newtheorem{definition}[theorem]{Definition}
\newtheorem{note}[theorem]{Note}

\newtheorem{remark}[theorem]{Remark}

\newcommand{\R}{\mathbb{R}}
\newcommand{\Prob}{\mathbb{P}}
\renewcommand{\Pr}{\Prob}
\renewcommand{\bf}{\bfseries}

\newcommand{\Q}{\mathbb{Q}}
\newcommand{\E}{\mathbb{E}}

\newcommand{\F}{\mathcal{F}}
\newcommand{\Fc}{\mathcal{F}}

\newcommand{\half}{\frac{1}{2}}

\newcommand{\pd}[2]{\frac{\partial #1}{\partial #2}}

\newcommand{\indic}[1]{\boldsymbol{1}_{\{\ensuremath{#1}\}}}

\newcommand{\ol}[1]{\overline{#1}}
\newcommand{\me}{\mathrm{e}}

\newcommand{\ie}{i.e.}
\newcommand{\eg}{e.g.}

\newcommand{\di}{\mathrm{d}}

\newcommand{\dy}{\di y}
\newcommand{\ds}{\di s}
\newcommand{\dt}{\di t}

\newcommand{\td}{\tilde}

\newcommand{\cts}{c}
\newcommand{\pits}{\pi}
\newcommand{\rts}{R}
\newcommand{\ra}{\gamma}

\newcommand{\dd}{\mathrm{d}}

\newcommand{\sA}{\mathcal A}
\newcommand{\sharpe}{\theta}
\newcommand{\Timefnct}{\beta}
\newcommand{\classC}{\mathcal{C}}

\begin{document}
\title{Utility theory front to back -- inferring utility from agents'
  choices}
\author{Alexander~M.~G.~Cox\thanks{e-mail:
        \texttt{A.M.G.Cox@bath.ac.uk};
        web: \texttt{www.maths.bath.ac.uk/$\sim$mapamgc/}}\\
        Dept.\ of Mathematical Sciences\\
        University of Bath\\
        Bath BA2 7AY
 \and David Hobson\thanks{e-mail:
        \texttt{D.Hobson@warwick.ac.uk};
        web: \texttt{www.warwick.ac.uk/go/dhobson/}}\\
        Department of Statistics\\
        University of Warwick\\
        Coventry CV4 7AL
 \and Jan Ob\l \'oj\thanks{e-mail:
        \texttt{obloj@maths.ox.ac.uk}; web:
        \texttt{www.maths.ox.ac.uk/$\sim$obloj/}}\\
        Mathematical Institute \emph{and}\\
	Oxford--Man Institute of Quantitative Finance\\
	University of Oxford\\
        Oxford OX1 3LB
}
\date{February 2012}
\maketitle
\begin{abstract}
  We pursue an inverse approach to utility theory and consumption \&
  investment problems. Instead of specifying an agent's utility
  function and deriving her actions, we assume we observe her actions
  (i.e.\ her consumption and investment strategies) and ask if it is
  possible to derive a utility function for which the observed
  behaviour is optimal. We work in continuous time both in a
  deterministic and stochastic setting. In the deterministic setup, we
  find that there are infinitely many utility functions generating a
  given consumption pattern. In the stochastic setting of the
  Black-Scholes complete market it turns out that the consumption and
  investment strategies have to satisfy a consistency condition (PDE)
  if they are to come from a classical utility maximisation
  problem. We show further that important characteristics of the agent
  such as her attitude towards risk (e.g.\ DARA) can be deduced
  directly from her consumption/investment choices.
\end{abstract}

\section{Introduction}

The study of investment and consumption problems in finance has a long
history, and there is large literature relating to these problems. In
general, however, the set-up and solution of the problems take the
following form: specify a utility function which describes the
investor's `desire' for wealth/consumption, and then solve a
stochastic optimisation problem to find the optimal investment and
consumption behaviour. Unfortunately, although we can postulate a
simple parametric form for the utility function, and hope to deduce
correspondingly simple forms for the optimal consumption/investment
strategies, it is
difficult to justify any claim that such a utility function accurately
represents the preferences of the agent. Moreover, attempts to elicit
utility functions directly are notoriously difficult, and prone to
paradoxes and inconsistencies.

In this work we approach consumption/investment problems from a
different, and possibly more natural, perspective.
Rather than supposing that we have previously divined an investor's
utility function, we suppose that we know their future consumption and
investment patterns, and ask whether we can compute a corresponding
utility function from the given behaviour. We believe that there are a
number of reasons why this is a natural question to ask:
\begin{itemize}
\item consumption and investment strategies are `observables' in that
  they can actually be measured from investors' actions,
  and therefore they are a more natural concept around which to build
  a model than the intangible utility function;
\item the framework will allow us to see how natural behaviour
  patterns in the consumption/investment setting relate to properties
  of the underlying utility function;
\item the analysis mirrors the robust approach to pricing and hedging
  (cf.~ \cite{CoxObloj:10,HobsonSurvey}) where one takes the
  vanilla option prices as observables and attempts to infer
  information about the prices of exotics and the dynamics of the
  price process of the underlying.
\end{itemize}
Our general question {regarding} how much {information} about {an}
agent's preferences and optimality criteria we can recover from her
behaviour and choices falls under the heading of \emph{revealed
  preferences} in Economics. It dates back (at least) to
\cite{Samuelson:48} and is sufficiently central and important that it
deserves an entry in the New Palgrave Dictionary of Economics
(\cite{Richter:08}). {Other related work in the economics literature
  includes \cite{GreenSrivastava:85}, who consider when a given
  consumption may be optimal for a utility maximising investor in a
  one-period, finite state model, and \cite{Mas-Colell:77}, where the
  observed quantities are the demand functions of consumers, and the aim
  is to recover the consumers' preferences.} In the financial
literature a similar ``reversed'' point of view was adopted by
\cite{DybvigRogers:97} who considered the recovery of an agent's
utility function from a single realisation of her consumption path,
working under the (strong) assumption of time homogeneity of agent's
utility function.

The closest to our work are papers of \cite{Black:68},
\cite{CoxLeland:00} and \cite{HeHuang:94}\footnote{We thank Thaleia
  Zariphopoulou who indicated these valuable references to us when
  this article neared completion. The manuscript \cite{Black:68} was
  published in a modified form as \cite{Black:88}, \cite{CoxLeland:00}
  was circulated informally in the 1980s, see Editor's note
  therein.}. These three papers considered the `inverse Merton
problem' on a finite time horizon while we consider the infinite
horizon case. However the results are close in spirit.  One of our
main results in the continuous time Black-Scholes market setting is
that consumption/investment strategies are compatible with a utility
maximisation framework if and only if they satisfy a certain PDE. We
call this Black's PDE as it was first derived in \cite{Black:68}. It
was then re-derived using discrete time arguments in
\cite{CoxLeland:00}. The analysis was then extended and made rigorous
in a remarkable paper of \citet{HeHuang:94}.  The key novelties of our
paper are firstly in considering an infinite horizon, which requires
dealing with the budget constraint and finiteness of the value
function, and secondly in offering a unified, mathematically rigorous
treatment of the problem. Moreover, we give several examples, and we show 
that they satisfy the sufficiency conditions of the theorem. This seems very 
difficult in the set-up of the pre-exisiting literature.

We make more detailed comments about the relationship between our work and 
that of \cite{Black:68},
\cite{CoxLeland:00} and \cite{HeHuang:94} in Remark
\ref{rk:BlackHH} below.  We note also that the main result linking an
agent's action via a PDE is similar in spirit to results in
\cite{Wang:93}. However in \cite{Wang:93} a full equilibrium model for
a representative agent is considered and we have a partial equilibrium
for a price taking agent.

The paper is organised as follows. In the first part of the paper,
Section \ref{sec:deterministic}, we work in a deterministic
setup. After {the} problem setup and a heuristic derivation of the
solution, the main theorem is given in Section \ref{sec:maindet}. In
Section \ref{sec:dara} we analyse what can be deduced from agents'
consumption about their risk attitudes and present explicit
examples. Finally, Section \ref{sec:admis_ut} comments on our
assumptions and the resulting restrictions.
As well as being interesting in its own right, this section
brings insight to the stochastic problem, by showing what we might
expect or not expect to be able to recover about $u$. Nonetheless, there
are also fundamental differences.

In the second part of the paper, Sections \ref{sec:stochastic} and 
\ref{sec:consequences}, we work
in {the} stochastic setting of {a} Black-Scholes market. Sections
\ref{sec:heuristics_primal} and \ref{sec:heuristics_dual} give a
heuristic derivation of the main result using the primal and the dual
approach respectively. The main theorem and its proof are then given
in Section \ref{sec:main_stoch}. 

Section \ref{sec:consequences} contains a 
discussion of some consequences of the main theorem and an extension.
Section \ref{sec:model_uncert} asks what happens if the parameters
of the Black-Scholes financial market are not known, and in particular asks if 
it is possible to recover the volatility and Sharpe ratio from the actions of 
the agent.
Section \ref{sec:riskaversion}
discusses the implication for reading off the risk attitudes of agents
from their actions. Section \ref{sec:timehom} focuses on the case
of time-homogenous strategies and presents explicit examples 
where consumption and investment are not linear in wealth. 
Finally, Section~\ref{sec:bdd} discusses a relaxation of the assumptions of 
the main theorem 
and in particular includes the case when an agent's consumption and 
possibly his wealth are uniformly bounded. Two examples are presented. 

Section \ref{sec:further} presents possible extensions of our work 
and future challenges.

\medskip

\noindent {\bf Notation.} We make the following notational
assumptions: throughout, an optimal consumption strategy will be
denoted by $\cts(t,w)$, where $t$ is the current time, and $w$ the
investors wealth at time $t$. Similarly, an optimal investment
strategy (in terms of the cash amount invested in a risky asset) will
be $\pits(t,w)$. A general consumption and investment process will be
$(C_t,\Pi_t)$. All stochastic processes will be denoted by capital
letters. Partial derivatives will be written $c_w(t,w)$ and
$c_t(t,w)$. There should be no confusion over subscripts $t$ since
applied to a (upper case) process it refers to a time parameter, and
applied to a (lower case) function it is a derivative.

\section{Deterministic setting}\label{sec:deterministic}

\subsection{Problem set-up}
We begin by considering the case where there is no stochastic
investment opportunity, so that we only observe the investor's
consumption over an infinite horizon. More specifically, suppose we
know that the investor who has wealth $w$ at time $t$ will consume an
amount $\cts(t,w) \dt$ in the time interval $t,t+\dt$, where
$\cts(t,w) \ge 0$, and suppose that we work in a situation with no
interest on savings (or equivalently,
all amounts are discounted back to their time-zero values). Then an
investor with initial wealth $x \ge 0$ will have time-$t$ wealth
$w(t,x)$ described by:
\begin{equation} \label{eq:determwdefn}
  \begin{split}
    w_t(t,x) & =  -\cts(t,w(t,x)) \\
    w(0,x) & = x.
  \end{split}
\end{equation}
Further, we impose that the budget constraint $w(t,x) \ge 0$ holds for
all $x$ and $t$, or, in terms of $\cts(t,w)$, that:
\begin{equation*}
  \int_0^\infty \cts(t,w(t,x))\dt \le x.
\end{equation*}

Our main concern is then the following. Suppose $\cts(t,w)$ is as
above, and suppose we are told $\cts(t,w(t,x))$ is optimal for the
problem:
\begin{equation}
  \label{eq:determprobdefn}
  v(x) = \sup_{\substack{C_t \ge 0, \\ \int_0^\infty C_t \dt \le x}}
  \int_0^\infty u(t,C_t)\, \dt,
\end{equation}
where the supremum is taken over processes $(C_t)_{t \ge 0}$.  What
can we infer about the function $u$?

\subsection{Heuristics}\label{sec:det_heuristics}
By introducing a Lagrangian term into \eqref{eq:determprobdefn} we
get:
\begin{eqnarray*}
  v(x) & = & \inf_{\lambda \ge 0} \sup_{C_t \ge 0} \left[
    \int_0^\infty u(t,C_t) \, \dt - \lambda \left( \int_0^\infty C_t
      \, \dt -x\right) \right]\\
  & = & \inf_{\lambda \ge 0} \sup_{C_t \ge 0} \left[ \int_0^\infty
    (u(t,C_t) - \lambda(x) C_t) \, \dt + x \lambda(x) \right].
\end{eqnarray*}
In the second line we write $\lambda = \lambda(x)$ to emphasise that
$\lambda$ will depend on the initial wealth.

Hence, for the optimal $C_t$, we get (supposing that $u$ is suitably
differentiable):
\begin{equation} \label{eq:determudash} u_c(t,\cts(t,w(t,x))) =
  \lambda(x),
\end{equation}
where the optimality of $\lambda$ implies $v_x(x) = \lambda(x)$.
Moreover, if we differentiate \eqref{eq:determudash} again, we get:
\begin{equation}
  \label{eq:determuddash}
  u_{cc}(t,\cts(t,w(t,x))) = -\frac{D(x)}{\frac{\partial}{\partial x}
    \left[ \cts(t,w(t,x)) \right]},
\end{equation}
where $D(x) = -\lambda_x(x) = -v_{xx}(x)$.  To then find
$u_{cc}(t,c)$, we need to assume that we can recover $x$ as a function
of $c$ and $t$. This will be the case if we assume that $\cts(t,w)$ is
increasing as a function of $w$. It seems to be a fairly natural
assumption to make in terms of investor behaviour, although note that
the assumption does then imply $|w(t,x_0)- w(t,x_1)|$ is decreasing in
$t$ --- that is, the `wealth paths' corresponding to different initial
wealths are `getting closer together' as time increases. Moreover, one
could imagine paths corresponding to two different initial wealths
merge at some later point. To rule out such behaviour we will also
impose that $\pd{}{x}\cts(t,w(t,x)) > 0$. Note as well that if this is
combined with the assumption that $D(x) > 0$ (or equivalently that the
value function is concave in $x$) we will have $u$ concave --- or a
decreasing marginal utility of additional consumption. Since these all
seem fairly plausible economic assumptions, we will work from this
point on under these assumptions.

Finally, observe that $u$ will necessarily be undetermined at least up
to addition of a function of the form $A(t) + B c$ and we would not
expect to fully recover $u$ from \eqref{eq:determuddash}.

\begin{example} \label{ex:CRRA} {\bf CRRA:} Suppose the optimal
  consumption strategy is: $\cts(t,w) = \kappa w$, for some $\kappa
  >0$ --- so the investor consumes a constant proportion of her
  wealth, and that she always consumes all of her wealth. Then it
  follows that $w(t,x) = x \me^{-\kappa t}$ and $\cts(t,w(t,x)) =
  \kappa x \me^{-\kappa t}$, and we can invert this to see that if she
  is consuming $c$ at time $t$, then her initial wealth was
  $\frac{c}{\kappa} \me^{\kappa t}$. Hence we get:
  \begin{equation*}
    u_{cc}(s,c) = -\frac{D\left( \frac{c}{\kappa} \me^{\kappa
          t}\right)}{\kappa \me^{-\kappa t}}.
  \end{equation*}
  Motivated by our knowledge of the form of the solution in the CRRA
  case, suppose we assume further that $D(x) = -v_{xx}(x) = \ra
  x^{-\ra-1}$
  for some positive $\ra$.  Then
  \begin{eqnarray*}
    u_{cc}(t,c) & = & \frac{-\ra\left( \frac{c}{\kappa} \me^{\kappa
          t}\right)^{-\ra-1}}{\kappa \me^{-\kappa t}} = -\ra \kappa^\ra \me^{-\ra\kappa t} c^{-\ra-1}
  \end{eqnarray*}
  Integrating this expression in $c$, we get:
  \begin{eqnarray*}
    u_c(t,c) & = & c^{-\ra} \kappa^\ra \me^{-\ra \kappa t} + \beta(t) = \left( \kappa x \me^{-\kappa t}\right)^{-\ra} \kappa^\ra \me^{-\ra  \kappa t} + \beta(t)\\
    & = & x^{-\ra} + \beta(t),
  \end{eqnarray*}
  but by \eqref{eq:determudash}, we know this expression must be
  independent of $t$, \ie{} $\beta(t) \equiv \beta$, and integrating
  once more, we get
  (provided $\ra \neq 1$):
  \begin{equation*}
    u(t,c) = \frac{c^{-\ra+1} \kappa^\ra \me^{-\ra \kappa t}}{1-\ra} + A(t)
    +\beta c
  \end{equation*}
  where $A$ is an unknown function of time, and $\beta$ is a
  constant. Note that these will not affect the choice of the optimal
  strategies derived from the utility function (assuming that
  $\int_0^\infty A(t) \, \dt$ is finite).
\end{example}

{We remark} that in the above example, we could have chosen any
positive function $D(\cdot)$, and we would end up with the
corresponding value function at time 0 given by $v_{xx}(x) = -D(x)$,
with exactly the same optimal consumption paths. This suggests we can
interpret the paths $(t,\cts(t,w(t,x)))$ as the contours where the
gradient of $u$ is constant, while the function $D$ encodes our
relative valuation of the different paths. Knowledge of consumption
paths does not reveal the relative valuations of the different paths
since there is no natural way of comparing the path with initial
wealth $x$ and the path with initial wealth $y$, simply from the
specification of the optimal paths. Specifying the function
$D(\cdot)$, however, does give an indication as to the relative
valuation of the different paths, and in order to recover $u(t,c)$, we
would expect to need to specify this function. We come back to this
issue below in Section \ref{sec:dara} and Example \ref{ex:Gex}.
Parallels in the setting where a risky asset is traded and {an} agent
{also} has to specify her investment strategy are drawn in Remark
\ref{rk:det_stoch} in Section \ref{sec:stochastic}.

\subsection{Main results}\label{sec:maindet}
Before we transform the above remarks into a theorem, we also note
that there may be a `maximal' solution to \eqref{eq:determwdefn},
given by $\bar{w}(t) = \sup_{x \ge 0} w(t,x)$ which may be finite for
$t>0$. In such a case, there is a `maximal' wealth path which comes
down from infinity in finite time, and since we assume we only see
behaviour from time zero, we will not observe any behaviour at higher
wealths, and therefore at higher consumptions than $\bar{c}(t) =
\sup_{x \ge 0} \cts(t,w(t,x))$. Thus we will not be able to infer
features of $u$ for
levels of consumption above $\bar{c}(t)$. Some thought confirms that
$\bar{c}(t)$ may be finite even if $\bar{w}(t)$ is equal to infinity
for all $t$. Mathematically, we will represent this fact by assuming
the function $u(t,c)$ is constant above $\bar{c}(t)$, but note that
there may be other possible choices of $u$ which produce the same
optimal choice of $\cts$ and $w$.

\begin{theorem} \label{thm:determmain} Suppose we are given functions
  $\{\cts(t,w) : w \in \R_+, t \ge 0\}$ such that $\cts(t,0) \equiv
  0$, $\cts(t,w)$ is locally Lipschitz continuous and strictly
  increasing in
  $w$. Let $w(t,x)$ be the (unique) solution to:\\
  \parbox{10cm}{
    \begin{eqnarray*}
      w_t(t,x) & = & -\cts(t,w(t,x)) \\
      w(0,x) & = & x,
    \end{eqnarray*}}
  \hfill
  \parbox{1cm}{\begin{eqnarray} \label{eq:wdefn} \end{eqnarray}}\\
  and suppose that
  \begin{equation*}
    \int_0^\infty \cts(t,w(t,x)) {\dt} = x,
  \end{equation*}
  and the function $\pd{}{x}\cts(t,w(t,x))$ exists and is strictly
  positive. Then there exists a function $u(t,c)$ such that $u_c(t,c)
  \ge 0$ and $u_{cc}(t,c) \le 0$, for which the problem:
  \begin{equation}
    \label{eq:determvfdefn}
    v(x) = \sup_{\substack{C_t \ge 0 : \\ \int_0^\infty C_t \, \dt \le
        x}} \int_0^\infty u(t,C_t)\, \dt
  \end{equation}
  is uniquely solved by the choice of consumption: $C_t =
  \cts(t,w(t,x))$ for each $x \ge 0$.
\end{theorem}

\begin{remark} In fact, as we shall see, there is a family of
  solutions $u$ for which the choice $C_t = \cts(t,w(t,x))$ is
  optimal. It should also be clear from the proof that Theorem
  \ref{thm:determmain} could be modified into an if and only if
  statement, albeit more technical and complicated than the current
  version.
\end{remark}

\begin{proof}
  Define $\bar{c}(t) = \sup_{x \ge 0} \cts(t,w(t,x))$, then for $0 \le
  c < \bar{c}(t)$, we can find a unique $x$ such that $c =
  \cts(t,w(t,x))$. Write this as $y(t,c)$, and note therefore that
  $y(t,\cts(t,w(t,x)))=x$, and $y(t,\bar{c}(t)) = \infty$. Also, by
  the assumption that $\pd{}{x}\cts(t,w(t,x))$ exists and is strictly
  positive, $y(t,c)$ is a differentiable function of $c$ with
  derivative
  \begin{equation*}
    y_c(t,c) = \frac{1}{\left.\pd{}{x}\cts(t,w(t,x))\right|_{x=y(t,c)}} .
  \end{equation*}
  Let $D(x)$ be a strictly positive function satisfying
  \begin{equation}
    \label{eq:determDcond}
    \int_x^\infty D(y) \, \dy < \infty,\quad \textrm{for every }x > 0.
  \end{equation}
  Then we can define a function $u$ by:
  \begin{equation} \label{eq:determudefn} u_c(t,c) =
    \begin{cases}
      \int_c^{\bar{c}(t)} \frac{D(y(t,\kappa))}{\left.\pd{}{x}
          \cts(t,w(t,x))\right|_{x=y(t,\kappa)}}\, \di\kappa &: c \le \bar{c}(t) \\
      0 &: c > \bar{c}(t) 
    \end{cases},
  \end{equation}
  where \eqref{eq:determDcond} ensures that the integral is finite for
  $c > 0$. Indeed, using the substitution $\xi = y(t,\kappa)$, we get:
  \begin{eqnarray}
    u_c(t,c) & = & \int_c^{\bar{c}(t)} \frac{D(y(t,\kappa))}{\left.\pd{}{x}
        \cts(t,w(t,x))\right|_{x=y(t,\kappa)}}\, \di\kappa
    \nonumber \\
    & = & \int_{y(t,c)}^{y(t,\bar{c}(t)) = \infty} \frac{D(\xi)}{\left.\pd{}{x}
        \cts(t,w(t,x))\right|_{x=\xi}}
    \left. \pd{}{x}\cts(t,w(t,x))\right|_{x=\xi} \,
    \di\xi \nonumber\\
    & = & \int_{y(t,c)}^{\infty} D(\xi) \, \di\xi. \label{eq:udashofc}
  \end{eqnarray}
  Then $u_c(t,c) \ge 0$ and $u_{cc}(t,c) \le 0$ so that
  $u(t,c)$ is strictly concave in $c$.  Also, writing $c=\cts(t,w(t,x))$ we
  find
  \begin{equation}
    u_c(t,\cts(t,w(t,x))) = \int_x^{\infty} D(\xi) \, \di\xi. \label{eq:determudasheq}
  \end{equation}

  Now we consider a general consumption path $C_t$ satisfying
  $\int_0^\infty C_t \dt \le x$. Then, using the concavity of
  $u(t,\cdot)$ and \eqref{eq:determudasheq}, we conclude:
  \begin{eqnarray}
\lefteqn{    \int_0^\infty \left[ u\left(t,C_t\right) -
u(t,\cts(t,w(t,x)))
\right] \, \dt} \label{eq:ucopt} \\
& \le & \int_0^\infty u_c(t,\cts(t,w(t,x))) (C_t -
    \cts(t,w(t,x))) \, \dt \nonumber \\
    & = & \left( \int_x^{\infty} D(\xi) \, \di\xi \right)
    \int_0^\infty (C_t - \cts(t,w(t,x))) \, \dt \nonumber\\
    & \le & 0 \nonumber
  \end{eqnarray}
  where the budget constraint gives the final step. Hence the given
  $\cts(t,w(t,x))$ is the optimal path as required. Finally, the
  inequality in \eqref{eq:ucopt} is strict, since $D(x)$ is strictly
  positive, unless $\cts(t,w(t,x)) = C_t$; hence $\cts(t,w(t,x))$ is
  also the unique optimal solution.
\end{proof}

\subsection{Inferring risk aversion from optimal consumption}\label{sec:dara}
So far, we have discussed the derivation of a utility function from an
initial choice of consumption behaviour. Can we extend this, and say
something about some other classical methods of describing investor
behaviour? For example, a natural question in this direction would be:
given a set of consumption paths can we determine whether the investor
has decreasing absolute/relative risk aversion?

As already observed in Section \ref{sec:det_heuristics} above, it
turns out that specifying the consumption paths alone is not
sufficient. We present examples below of two utility functions, one
with decreasing absolute risk aversion and one with increasing
absolute risk aversion, which yield the same optimal consumptions
paths. In essence, consumption alone does not tell us how the investor
compares different wealths. This is specified by the additional
function $D$. We can think of $D(x)$ (or more accurately
$\int_x^\infty D(y) \, \dy$) as determining the relative weightings of
different initial wealths: when $D(x)$ is large, the additional
utility of an agent from a small increase in initial wealth above $x$
is large, when $D(x)$ is small, the additional utility is also small.
In what follows, we say that an agent with consumption paths $c(t,w)$
has \emph{relative weighting of initial wealths $D(x)$} if $D(x)$ is
differentiable, satisfies \eqref{eq:determDcond}, and {the} agent's
utility is specified via \eqref{eq:determudasheq}.

We start with a simple observation about the role of the function $D$.
\begin{note}
  The {\it Inada condition} --- that is, that for all $t$, $u_c(t,c)$
  takes all values in $[0,\infty)$, is equivalent to $\int_x^\infty
  D(y) \dy \uparrow \infty$ as $x \downarrow 0$.
\end{note}
We {now} analyse in detail the risk aversion of the investor. We
concentrate on absolute risk aversion, but we observe that similar
{results} can be {derived} for relative risk aversion.

\begin{definition}\label{def:ra}
  For a utility function $u$, the absolute risk aversion is given by
  \begin{equation*}
    \rho(t,c) = -\frac{u_{cc}(t,c)}{u_c(t,c)}.
  \end{equation*}
  We say that an investor is DARA (decreasing absolute risk aversion)
  if $\rho_c(t,c) \le 0$ for all $t, c\ge 0$.  Similarly, we say an
  investor is CARA (constant absolute risk aversion) or IARA
  (increasing absolute risk aversion) if respectively $\rho_c(t,c) =
  0$ or $\rho_c(t,c) \ge 0$, for all $t,c\geq 0$.
\end{definition}

Recall that $u$ is recovered only up to an affine function. We should
note that our normalisation $u_c(t,\infty)=0$, or more precisely
$\lim_{x \to \infty} u_c(t,\cts(t,w(t,x))) = 0$, which is implicit in
the equation \eqref{eq:determudash} and explicit in
\eqref{eq:determudasheq}, and follows from the use of $\bar{c}(t)$ as
a reference point in \eqref{eq:determudefn}, has a consequence on the
value of the function $\rho(t,c)$. A different reference point might
change the absolute risk aversion.

\begin{prop} \label{thm:ARAsign} Suppose an investor has consumption
  paths $\cts(t,w)$ and relative weighting of initial wealths
  $D(x)$. Then the sign of $\rho_c(t,c)$ is the same as the sign of:
  \begin{equation*}
    \frac{D_x(x)}{D(x)} + \frac{D(x)}{\int_x^\infty D(y) \, \dy} -
    \frac{\pd{^2}{x^2}\cts(t,w(t,x))}{\pd{}{x}\cts(t,w(t,x))} \equiv
    \pd{}{x}
    \ln\left(\frac{D(x)}{\pd{}{x}\cts(t,w(t,x)) \int_x^\infty D(y) \,
        \dy} \right),
  \end{equation*}
  {evaluated at $x = y(t,c)$.}
\end{prop}

\begin{cor}
  An investor is DARA if and only if:
  \begin{equation} \label{eq:DARAequiv} \frac{D_x(x)}{D(x)} +
    \frac{D(x)}{\int_x^\infty D(y) \, \dy} \le \inf_{t \ge 0}
    \frac{\pd{^2}{x^2}\cts(t,w(t,x))}{\pd{}{x}\cts(t,w(t,x))},\quad
    x>0.
  \end{equation}
  An investor is CARA if and only if
  \begin{equation*}
    \frac{D_x(x)}{D(x)} + \frac{D(x)}{\int_x^\infty D(y) \, \dy}  =
    \frac{\pd{^2}{x^2}\cts(t,w(t,x))}{\pd{}{x}\cts(t,w(t,x))}
  \end{equation*}
  so that in particular, the right hand side of the equation is
  independent of $t$. Finally an investor is IARA if and only if:
  \begin{equation} \label{eq:IARAequiv} \frac{D_x(x)}{D(x)} +
    \frac{D(x)}{\int_x^\infty D(y) \, \dy} \ge \sup_{t \ge 0}
    \frac{\pd{^2}{x^2}\cts(t,w(t,x))}{\pd{}{x}\cts(t,w(t,x))}, \quad
    x>0.
  \end{equation}
\end{cor}

\begin{proof}[Proof of Proposition~\ref{thm:ARAsign}]
  It follows from \eqref{eq:determudasheq} that:
  \begin{equation*}
    \rho(t,\cts(t,w(t,x))) = \frac{D(x)}{\pd{}{x} \cts(t,w(t,x))
      \int_x^\infty D(y) \, \dy}.
  \end{equation*}
  Since $c(t,w(t,x))$ is increasing in $x$, $\rho(t,c)$ is increasing
  in $c$ if and only if the right-hand-side of the above expression is
  increasing in $x$, if and only if the logarithm of the
  right-hand-side is increasing in $x$.
\end{proof}

\begin{example}\label{ex:Gex}
  Consider the consumption function of Example~\ref{ex:CRRA}, so that
  $\cts(t,w) = \kappa w$ and $\cts(t,w(t,x)) = \kappa x \me^{-\kappa
    t}$. Then:
  \begin{equation*}
    \frac{\pd{^2}{x^2}\cts(t,w(t,x))}{\pd{}{x}\cts(t,w(t,x))} = 0.
  \end{equation*}
  If we consider a function $D(x) = \ra x^{-\ra-1}$ with
  $\ra >0$, then
  \begin{equation*}
    \frac{D_x(x)}{D(x)} + \frac{D(x)}{\int_x^\infty D(y) \, \dy} =
    -\frac{1}{x} < 0,
  \end{equation*}
  so the corresponding investor is DARA. On the other hand, for the
  choice $D(x) = x \me^{-\eta x^2}$ with $\eta >0$,
  \begin{equation*}
    \frac{D_x(x)}{D(x)} + \frac{D(x)}{\int_x^\infty D(y) \, \dy} =
    \frac{1}{x} > 0,
  \end{equation*}
  and the investor is IARA. The case $D(x) = \me^{-\zeta x}$ gives a
  CARA investor. Note that in the last two cases we
  necessarily have $u(t,\infty)<\infty$, whereas in the first case the
  finiteness of $u(t,\infty)$ depends on the sign of $(\ra -1)$.
\end{example}

\begin{example}
  The purpose of this example is to show that explicit answers may
  still be available beyond the CRRA case in which consumption is
  proportional to wealth. Again we find that knowledge of the
  consumption path alone is not sufficient to determine the attitude
  to risk.

  Suppose we have a concave, increasing function $G(z)$ of class
  $\classC^3$ and such that $G(0) = 0, G_z(0) = 1$ and $G(z)/z \to 0$
  as $z \to \infty$. Let $w(t,x) = \frac{1}{t}G(xt)$. Then it follows
  that $w(0,x) = x$ and:
  \begin{align*}
    \cts(t,w(t,x)) & = -\frac{\partial}{\partial t}\left[ \frac{1}{t} G(xt)\right]
    \\
    & = \frac{1}{t^2} \left[ G(xt) - xt G_z(xt)\right] = \frac{w}{t}
    -\frac{1}{t^2} G^{-1}(tw) G_z( G^{-1}(tw))
  \end{align*}
  which is positive by concavity. In particular, we get:
  \begin{equation*}
    \frac{\pd{^2}{x^2}\cts(t,w(t,x))}{\pd{}{x}\cts(t,w(t,x))} =
    \frac{1}{x} + t \frac{G_{zzz}(xt)}{G_{zz}(xt)}
  \end{equation*}

  One simple example of such a function is $G(z) = \ln(1+z)$, in which
  case we get $\cts(t,w) = \frac{1}{t^2}(tw + e^{-wt}-1)$ and:
  \begin{equation*}
    \frac{\pd{^2}{x^2}\cts(t,w(t,x))}{\pd{}{x}\cts(t,w(t,x))} =
    \frac{1}{x}- \frac{2t}{1+xt}.
  \end{equation*}
  This expression is decreasing in $t$, so we can conclude that
  \begin{equation*}
    \inf_{t \ge 0}
    \frac{\pd{^2}{x^2}\cts(t,w(t,x))}{\pd{}{x}\cts(t,w(t,x))}  =
    \lim_{t \to \infty}\left[ \frac{1}{x}- \frac{2t}{1+xt} \right]
    =  -\frac{1}{x}
  \end{equation*}
  and we see that the corresponding investor is DARA if we take $D(x)
  = \ra x^{-\ra-1}$ for $\ra > 0$. On the other hand,
  \begin{equation*}
    \sup_{t \ge 0}
    \frac{\pd{^2}{x^2}\cts(t,w(t,x))}{\pd{}{x}\cts(t,w(t,x))} =
    \lim_{t \to 0} \left[ \frac{1}{x}- \frac{2t}{1+xt}\right] \\
    =  \frac{1}{x}.
  \end{equation*}
  so the choice $D(x) = x \me^{-\eta x^2}$ for any $\eta >0$, gives an
  IARA investor.

  Another example arises by taking $G(z) = 1-\me^{-z}$. In this case,
  we have $\cts(t,w) = \frac{w}{t} +(1-wt) t^{-2} \ln(1-wt) $, and
  \begin{equation*}
    \frac{\pd{^2}{x^2}\cts(t,w(t,x))}{\pd{}{x}\cts(t,w(t,x))} =
    \frac{1}{x}- t.
  \end{equation*}
  As before, taking \eg{} $D(x) = x \me^{- \eta x^2}$, gives an IARA
  investor. However there is no choice of $D(x)$ for which the
  investor will be DARA. Note that in this example, since $G$ is
  bounded by 1, the investor's wealth will be below $\frac{1}{t}$ at
  time $t$, no matter
  how large their initial wealth.

\end{example}

\subsection{Admissible utility functions}\label{sec:admis_ut}
It is natural to ask if we can recover all utility functions $u$ (up
to addition of a function $A(t) + B c$) from the above setup? The
answer is no.

Consider for example functions of the form: $u(t,c) = U(c)$ for some
increasing concave function $U(\cdot)$. Such functions correspond to
optimal paths which are constant, but of course, these have infinite
total consumption. Agents with finite initial wealth will try to
spread the total consumption as evenly as possible across the whole
time horizon, but there will be no sensible `optimal' consumption.
There may also be cases when optimal consumptions exist but are not
covered by our framework. For example, one may construct utility
functions for which optimal consumption paths are zero for a while and
then leave zero to follow a positive path.

Our aim in Section \ref{sec:deterministic} was to
consider the extent to which knowledge of optimal consumption paths
can be used to determine the utility in the deterministic case. To
obtain a complete and coherent description we worked under plausible,
but not necessary, assumptions e.g.\ that consumption levels are
strictly increasing in current wealth. The key discovery is that in
the deterministic case there is no way to compare utilities across
different optimal consumption paths.  We shall see that this situation
is rather special, and that the picture is different in the stochastic
case.

\section{Stochastic Setting} \label{sec:stochastic}

We now turn to a more sophisticated version of the above problem, by
considering what happens when we add the possibility of investment in
a stochastic asset. Specifically, we suppose there is a risky asset
$P_t$, where $P_t$ is a Black-Scholes asset so that it has dynamics:
\begin{equation}\label{eq:price_dynamics}
  \frac{\di P_t}{P_t} = \sigma (\di B_t + \sharpe\, \dt) + r \, \dt .
\end{equation}
Here $\sigma$ is the asset volatility, $\sharpe >0$ is the Sharpe
ratio, and $r$ is the interest rate, which are all assumed to be
constant, and $B_t$ a standard Brownian motion. The investor now has
to choose a rate of consumption $C_t$ and also an amount, $\Pi_t$,
which is to be invested in the risky asset. Then her wealth at time
$t$, $W_t$, is the solution to:
\begin{equation}\label{eq:wealthequation}
  \di W_t = r W_t \dt - C_t \dt + \Pi_t \sigma (\di B_t + \sharpe \dt),
\end{equation}
subject to $W_0=x$. Where we wish to highlight the dependence on
initial wealth $x$ or strategy $(C,\Pi)$ we may write this as $W_t^{x,
  C, \Pi}$.

The investor will specify an optimal pair of investment and
consumption strategies, $\Pi_t$ and $C_t$ which attain the supremum
in:
\begin{equation}
  \label{eq:stocproblem}
  \sup_{C_t, \Pi_t : W_t^{C,\Pi} \ge 0} \E\left[ \int_0^\infty u(t,C_t) \, \dt \right]
\end{equation}
subject to a budget constraint $W_t \ge 0$ for all $t \ge 0$. Here $u$
is an unknown function which we aim to find.

As usual, the above generalises to an optimal control problem, which
has value function:
\begin{equation*}
  v(t,w) = \sup_{C_s, \Pi_s : W_s^{C,\Pi} \ge 0} \E \left[ \int_t^\infty
    u(s,C_s) \, \ds \big| W_t = w\right].
\end{equation*}
Standard theory tells us that for a general pair $(C_t,\Pi_t)$ the
process $M_t = \int_0^t u(s,C_s)\, ds + v(t,W_t)$ must be a
supermartingale, and under the optimal strategy will be a
martingale. Applying It\^o's Lemma to $M_t$, we see that the relevant drift
($\dt$) terms are:
\begin{equation} \label{eq:martcond1} u(t,C_t) + v_t(t,W_t) +
  v_w(t,W_t)\left[rW_t -C_t +\Pi_t \sigma \sharpe\right] + \half
  v_{ww}(t,W_t) \sigma^2 \Pi_t^2.
\end{equation}

We assume that the optimal strategy takes the form $(C_t = c(t,W_t),
\Pi_t = \pi(t,W_t)).$ Then, by analysing this equation, and
considering possible solution terms $v(t,w)$, we prove in Theorem
\ref{thm:Dual} that there is a function $u(t,c)$
for which the pair $(\cts(t,w),\pits(t,w))$ is optimal if and only if these functions satisfy:
\begin{equation}
  \label{eq:cpicond}
  \frac{\cts(t,w)}{\pits(t,w)} -\frac{r
    w}{\pits(t,w)} + \frac{\sigma^2}{2}\pits_w(t,w)
  + \int_{\cdot}^w \frac{\pits_t(t,\td{w})}{
    \pits(t,\td{w})^2} \, \di\td{w}
  = \Timefnct(t)
\end{equation}
for some function $\Timefnct(t)$ --- in particular, the left hand side
is independent of $w$.  This consistency relationship between $\pi$
and $c$ for them to be the solution of an optimal
consumption/investment problem of the type \eqref{eq:stocproblem} was
first derived by \citet{Black:68} (published later in a modified form
as \cite{Black:88}), then by \cite{CoxLeland:00} and subsequently
generalised and made rigorous by \citet{HeHuang:94}, see Remark
\ref{rk:BlackHH} below.  Before stating and proving our main result,
Theorem \ref{thm:Dual}, we give heuristic derivations of
\eqref{eq:cpicond} using both primal and dual approaches to
\eqref{eq:stocproblem}.

\subsection{Heuristics: the primal
  approach}\label{sec:heuristics_primal}

To motivate the condition in \eqref{eq:cpicond}, it turns out to be
instructive to look at a more general problem: we introduce a function
$\Psi(t,w)$ and then consider
\begin{equation}
  \label{eq:vfdefnmod}
  v(t,w) = \sup_{\Pi_s,C_s}
  \E \left[\int_t^{\infty} \left( u(s,C_s) + \Psi(s,W_s)\right) \, \ds \big|
    W_t=w\right].
\end{equation}

As before our starting point is an assumption that the optimal
strategy takes the form $(C_t = c(t,W_t), \Pi_t = \pi(t,W_t)).$ Then,
by deriving an expression for $\Psi$ in terms of the functions
$\pits(t,w), \cts(t,w)$, we will be able to recover the condition
\eqref{eq:cpicond} in the special case where $\Psi$ is a function of
time alone.

In the same way that we derived \eqref{eq:martcond1}, we can get the
martingale condition corresponding to \eqref{eq:vfdefnmod} which is
\begin{align}
  \sup_{C}& \left[u(t,C) - v_w(t,W_t)C\right] + \sup_{\Pi}\left[\Pi
    \sigma \sharpe v_w(t,W_t) + \half \sigma^2 \Pi^2
    v_{ww}(t,W_t)\right] \nonumber \\
  & {}+ v_t(t,W_t) + v_w(t,W_t) r W_t + \Psi(t,W_t) =
  0. \label{eq:martcond2}
\end{align}
In particular, the optimal choice of $\Pi$, namely $\pits(t,w)$,
should satisfy:
\begin{equation} \label{eq:pistar} \pits(t,w) = -\frac{\sharpe
    v_w(t,w)}{\sigma v_{ww}(t,w)},
\end{equation}
which in turn suggests we can write:
\begin{equation} \label{eq:vexpress} v_w(t,w) = \exp\left\{ A(t) -
    \int_\cdot^w \frac{\sharpe}{\sigma \pits(t,\td{w})} \,
    \di\td{w}\right\},
\end{equation}
where $A(t)$ is some function of $t$ to be specified. Our aim is now
to use this expression to remove terms involving the function $v$ from
\eqref{eq:martcond2}. To this end, it will be easier to consider the
derivative of \eqref{eq:martcond2}.

Suppose that $u$ is concave and differentiable in $c$ and introduce
the convex dual, $\tilde{u}(t,\xi) = \sup_{\chi} \left(u(t,\chi) - \xi
 \chi \right)$. Note that we have $\tilde{u}_\xi(t,\xi) = -\chi^*$,
where
$\chi^*$ is the choice of $\chi$ which attains the supremum.

Substituting the optimal actions $\cts(t,w)$ and $\pits(t,w)$ into
\eqref{eq:martcond2}, we get:
\begin{equation}
  \label{eq:martcond3}
  0 = \td{u}(t,v_w(t,w)) + \half \pits(t,w) \sigma \sharpe v_w(t,w) + v_t(t,w) +
  v_w(t,w) r w + \Psi(t,w),
\end{equation}
and differentiating \eqref{eq:martcond3} with respect to $w$, we
obtain:
\begin{align}
  0 = -v_{ww}(t,w)\cts(t,w) + & \half \pits(t,w) \sigma \sharpe
  v_{ww}(t,w) + \half \sigma \sharpe v_w(t,w) \pits_w(t,w) +
  v_{tw}(t,w) \nonumber\\ & {} + v_{ww}(t,w) r w + v_w(t,w) r +
  \Psi_w(t,w).
  \label{eq:martcond4}
\end{align}
If we now differentiate \eqref{eq:vexpress} in the time variable, we
see that we must have:
\begin{equation}
  \label{eq:vdotdash}
  v_{tw}(t,w) = \left[ A'(t) + \int_\cdot^w \frac{\sharpe}{\sigma
      \pits(t,\td{w})^2}\pits_t(t,\td{w})\, \di\td{w}
  \right] v_w(t,w),
\end{equation}
so that \eqref{eq:martcond4} becomes:
\begin{align*}
  0 = & \left[\half \pits(t,w) \sigma \sharpe + rw - \cts(t,w) \right]
  v_{ww}(t,w)
  + \Psi_w(t,w) \\
  &{} + \left[ \half \sigma \sharpe \pits_w(t,w) + r +
    A'(t) + \int_\cdot^w \frac{\sharpe}{\sigma
      \pits(t,\td{w})^2}\pits_t(t,\td{w})\, \di\td{w} \right]
  v_w(t,w).
\end{align*}
Finally, dividing through by $v_w(t,w)$ we have:
\begin{align*}
  - \frac{\Psi_w(t,w)}{v_w(t,w)} = & \left[\half \pits(t,w) \sigma
    \sharpe + rw - \cts(t,w) \right] \frac{v_{ww}(t,w)}{v_w(t,w)} +
  \half \sigma \sharpe \pits_w(t,w) \\ & {} + r + A'(t) +
  \int_\cdot^w \frac{\sharpe}{\sigma
    \pits(t,\td{w})^2}\pits_t(t,\td{w})\, \di\td{w},
\end{align*}
and using \eqref{eq:pistar}, we get:
\begin{align}
  \Psi_w(t,w) \exp&\left\{-A(t) + \int_\cdot^w \frac{\sharpe}{\sigma
      \pits(t,\td{w})} \, \di\td{w} \right\} \nonumber \\ & {}= \half
  \sharpe^2 + \frac{rw\sharpe}{\sigma \pits(t,w)} - \frac{\sharpe
    \cts(t,w)}{\pits(t,w) \sigma} -\half \sigma \sharpe \pits_w(t,w) -
  r - A'(t) \nonumber \\ & \quad \quad {}- \int_\cdot^w
  \frac{\sharpe}{\sigma \pits(t,\td{w})^2}\pits_t(t,\td{w})\,
  \di\td{w}.
  \label{eq:Bdash}
\end{align}
Since we have not yet fixed the constant of integration $A(t)$, we are
free to choose this. Because our main interest is in the case where
$\Psi(t,w)$ is independent of $w$, it follows that we are interested
in cases where we can make the expression on the right-hand side of
\eqref{eq:Bdash} disappear, which will occur whenever the expression
\begin{equation*}
  \frac{ \sharpe^2}{2} +
  \frac{r\sharpe w}{\sigma \pits(t,w)} - \frac{\sharpe
    \cts(t,w)}{\sigma\pits(t,w) } -\frac{ \sigma \sharpe}{2}
  \pits_w(t,w) - r - \int_1^w
  \frac{\sharpe}{\sigma \pits(t,\td{w})^2}\pits_t(t,\td{w})\,
  \di\td{w}
\end{equation*}
is independent of $w$.  Differentiating once more in $w$, and
rearranging, we see that this is equivalent to $\pits, \cts$
satisfying:
\begin{equation}
  \label{eq:cpiPDE}
  \pits_t(t,w) = -\frac{\sigma^2}{2}
  \pits(t,w)^2\pits_{ww}(t,w) + (\cts(t,w)-rw) \pits_w(t,w) -
  \pits(t,w)\cts_w(t,w) + r\pits(t,w).
\end{equation}
This is Black's equation (\citet[Equation (9)]{Black:68}, see also
\citet[Equation (47)]{CoxLeland:00} and
\citet[Equation (26)]{HeHuang:94}).  Equivalently, defining
$\rts(t,w):=\frac{\cts(t,w)}{\pits(t,w)}$, we have that $\rts$ solves:
\begin{equation}\label{eq:cpiPDE2}
  \rts_w(t,w)=\frac{r}{\pits(t,w)}-\frac{1}{\pits(t,w)^2}\left(\pits_t(t,w)+r
    w \pits_w(t,w)\right)-\frac{\sigma^2}{2}\pits_{ww}(t,w),
\end{equation}
and integrating we arrive at Black's PDE in integrated form
\eqref{eq:cpicond}
\begin{equation}
  \label{eqn:black}
  \int_1^w \frac{\pits_t(t,\xi)}{(\pits(t,\xi))^2} \di \xi
  +  \frac{\sigma^2}{2}
  \pits_w(t,w) + \frac{\cts(t,w)}{\pits(t,w)} - r \frac{w}{\pits(t,w)} =
  \Timefnct(t)
\end{equation}
for some function $\Timefnct(t)$, independent of $w$.

\subsection{Heuristics: the dual approach}\label{sec:heuristics_dual}
We now give a second derivation of Black's equation using a dual
approach to the consumption/investment problem. We will use this
approach to give our main theorem below.

For the problem \eqref{eq:stocproblem} we can rewrite the budget
constraint as
\[ \E \left[ \int_0^\infty C_t Z_t \dt \right] = x, \] where $(Z_t)_{t
  \geq 0}$ is the state-price density process and is given by
\[ Z_t = \exp \left( - r t - \sharpe B_t - \frac{\sharpe^2}{2} t
\right). \] With this formulation the problem becomes to find
\[ \sup_{C_t} \E \left[ \int_0^\infty u(t,C_t) \dt - \lambda \left(
    \int_0^\infty C_t Z_t \dt - x \right) \right]
\]
for an appropriate Lagrange multiplier $\lambda = \lambda(x)$.  This
expression is bounded by $\lambda x + \E[\int_0^\infty
\tilde{u}(t,\lambda Z_t) \dt] $ and for optimality we must have that
$u_c(t,C_t)=\lambda Z_t$ so that writing $I(t,\cdot)$ for the inverse
in space of $u_c(t,\cdot)$ we deduce that the optimal consumption
takes the form $C_t = I(t, \lambda Z_t)$.

Now assume that the optimal strategy is a given function $c=c(t,w)$ of
time and wealth so that $C_t = c(t,W_t)$. It follows that $W_t = f(t,
\lambda Z_t)$ for some $f=f(t,z)$ which depends on the (now unknown
$u$) through $f = c^{-1} \circ I$.

Then, by It\^{o}'s Lemma,
\begin{eqnarray*}
  \di W_t & =  & \lambda f_z(t, \lambda Z_t) \di Z_t + f_t(t, \lambda Z_t) \dt +
  (1/2) f_{zz}(t, \lambda Z_t) \lambda^2 \di \langle Z\rangle_t \\
  & = & - \sharpe \lambda Z_t f_z \di B_t + \{ f_t + (1/2) \sharpe^2
  \lambda^2 Z^2 f_{zz} - r \lambda Z_t f_z \} \dt.
\end{eqnarray*}
Comparing this with the wealth dynamics \eqref{eq:wealthequation} we
have
\begin{equation}
  \label{pidef}
  \sigma \pits(t,f(t,z)) = - \sharpe z f_z(t,z),
\end{equation}
\begin{equation}
  \label{Cdef}
  rf(t,z) - c(t,f(t,z)) + \sharpe \sigma \pits(t,f(t,z)) =
  f_t(t,z) + (1/2) \sharpe^2 z^2 f_{zz}(t,z) - r z f_z(t,z).
\end{equation}
Then $\sigma \pits_w f_z = - \sharpe f_z - \sharpe z f_{zz}$ and
$\sigma \pits_t + \sigma \pits_w f_t = - \sharpe z f_{tz}$ so that
\begin{equation}
  \label{w''eqn}
  \sharpe^2 z^2 f_{zz} = (\sharpe + \sigma \pits_w)\sigma \pits,
\end{equation}
\begin{equation}
  \label{wdoteqn}
  f_{tz}/f_z = \pits_t/\pits + \pits_w f_t /\pits .
\end{equation}
Putting (\ref{w''eqn}) into (\ref{Cdef}) gives
\begin{equation}
  \label{maineqn}
  rf - \cts + \sigma \sharpe \pits = \frac{\sigma}{2} \pits (\sharpe +
  \sigma \pits_w) + f_t + \frac{r \sigma}{\sharpe} \pits .
\end{equation}
Differentiating with respect to $z$, dividing by $f_z = - \sigma
\pits/(\sharpe z)$, using (\ref{wdoteqn}) and (\ref{maineqn}) to
eliminate $f_{tz}$ and $f_t$ and multiplying by $\pits$ we finally get
\begin{equation}
  \label{picfundeqn1}
  \begin{split}
  (r - \cts_{w}(t,w)) \pits(t,w) - \frac{\sigma^2}{2}&(\pits(t,w))^2
  \pits_{ww}(t,w) - \pits_t(t,w) \\& - r \pits_w(t,w) w + \pits_w(t,w) c(t,w) = 0,
\end{split}
\end{equation}
which is Black's PDE \eqref{eq:cpiPDE}.

  \begin{remark}
    Our motivation so far has been the following: we have supposed
    that both the consumption and investment functions have been
    stated for all times and wealths, and we have derived the
    consistency condition \eqref{eqn:black} as a necessary condition
    that these functions must satisfy. However, the above calculations
    also suggest an alternative way of viewing the setup. Suppose
    instead our agent specifies her consumption (at all times and
    wealths), and her initial investment strategies at all wealths
    (\ie{} $\left\{\pi(0,w)\right\}_{w \ge 0}$). Then, under the
    assumption that the agent is a utility maximiser, we can solve the
    parabolic PDE \eqref{picfundeqn1} to deduce $\pi(t,w)$ at times $t
    \ge 0$. Note that the utility function itself is bypassed in the
    sense that we do not need to specify it to deduce $\pi(t,w)$. This
    was one of the motivating observations for \cite{Black:68}.
  \end{remark}

\subsection{Main results}\label{sec:main_stoch}
Given a pair of processes $(C,\Pi) \equiv (C_s, \Pi_s)_{s \geq 0}$
define the associated wealth process $(W^{x,C,\Pi}_s)_{s \geq 0}$ for
initial wealth $x$ by
\begin{equation}
  \label{Wdef}
  W^{x,C,\Pi}_s = x + \int_0^s \Pi_u \sigma(\di B_u + \sharpe \di u) + \int_0^s
  (rW^{x,C,\Pi}_u - C_u) \di u.
\end{equation}
We say that $C,\Pi$ is admissible if \eqref{Wdef} admits a strong
  solution $W^{x,C,\Pi}_s $ with $W^{x,C,\Pi}_s \geq 0$ for all $s$
and we write $\sA = \sA(x)$ for the space of admissible strategies.
Note that if $C,\Pi$ is admissible then, writing $W$ for
$W^{x,C,\Pi}$,
\begin{equation} \label{eq:ZWdynamics}
\di (Z_sW_s) = Z_s(\sigma \Pi_s -
  \sharpe W_s) \di B_s - Z_s C_s \ds .
\end{equation}
Hence
$(Z_s W_s)_{s \geq 0}$ is a non-negative supermartingale so that if
$W_s=0$ then, for $t \geq s$, $\E[W_t Z_t|\F_s] \leq W_s Z_s=0$ and
hence $W_t=0$ almost surely. Thus zero is absorbing for any admissible
strategy.

We suppose we are given functions $\cts=\cts(t,w)$ and $\pits =
\pits(t,w)$ and we aim to find, where possible, $u$ such that $C_t =
\cts(t,W^{x,C,\Pi}_t), \Pi_t = \pits(t,W^{x,C,\Pi}_t)$ is optimal for
\eqref{eq:stocproblem}. We start by defining the class of utility
functions {we consider} and imposing further
assumptions on our inputs. We focus here on the ``regular" case which yields a clean simple statement of the main result. Possible extensions which relax the assumptions on $u, \pits$ and $\cts$ are discussed in Section \ref{sec:bdd}.

Recall that a function $\phi(t,x)$ is {\it locally H\"older continuous}
on a set
$D$ if, for every $(t,x) \in D$, there is some neighbourhood $U$ of
$(t,x)$, and some $\alpha \in (0,1]$ such that
\begin{equation*}
  \sup_{(t',x'),(t,x) \in U}
  \frac{|\phi(t,x)-\phi(t',x')|^2}{\left(|x-x'|^2 +
|t-t'|\right)^\alpha}
  < \infty.
\end{equation*}
Note that a function which is locally-H\"older continuous is jointly
continuous.

\begin{definition}\label{def:utility}
  We say that a function $u:[0,\infty)^2\to [-\infty,\infty)$ is a
  \emph{regular utility function} if for any $t\geq 0$, $u(t,\cdot)$
  is twice continuously differentiable, strictly concave and strictly
  increasing, and satisfies the Inada condition: $u_c(t,0)=\infty$ and
  $u_c(t,\infty)=0$. Further, $I(t,\cdot)$ defined to be the inverse
  in space of $u_c$ (so that $u_c(t,I(t,z))=z$) is such that $I_z$ is
  locally H\"older continuous on $(0,\infty)^2$.
\end{definition}
If a utility function $u$ is given we denote the set
of admissible strategies for which the reward in
\eqref{eq:stocproblem} is well defined by $\sA^u(x)=\{(C,\Pi)\in
\sA(x): \E\int_0^\infty u(t,C_t)^+\dt<\infty\textrm{ or
}\E\int_0^\infty u(t,C_t)^-\dt<\infty\}$.

\begin{definition}\label{def:regularpic}
We say that $(c,\pi)$ is a \emph{regular consumption/investment
pair} if
\begin{itemize}
\item for each $t\geq 0$, $\cts(t,0)=0$ and $\cts(t,\cdot)$ is
  strictly increasing, unbounded and differentiable with $c_w(t,w)$ locally
  H\"older continuous on $(0,\infty)^2$.
\item for each $t\geq 0$, $\pits(t,0)=0$, $\pits(t,\cdot)$ is strictly
  positive and $\int_{0}^1 \di \xi / \pi(t,\xi)= \infty =
  \int_1^{\infty} \di \xi / \pi(t,\xi)$. Further, $\pits=\pits(t,w)$
  is continuously differentiable in both arguments on $(0,\infty)^2$.
\end{itemize}
Finally, $c$, $\pi$ are such that the SDE
  \begin{equation}\label{eq:wealth_dynamics}
    \di W^x_t =  \pits(t,W^x_t) \sigma(\di B_t + \sharpe \dt) +
    (rW^x_t - \cts(t,W^x_t)) \dt,
    \hspace{10mm}  W^x_0=x,
  \end{equation}
  has a strong solution.
\end{definition}
When we want to emphasize the dependence on $c$ and $\pi$ we denote
the solution to \eqref{eq:wealth_dynamics} by
$W^{x}=W^{x,\cts,\pits}$.

Assuming that $(c,\pi)$ is a regular consumption/investment pair
define $Y(t,c)$ to be the inverse to $\cts(t,w)$ so that
$Y(t,\cts(t,w))=w$.  Suppose further that $\cts,\pits$ satisfy
\eqref{eqn:black} and let \( A(t) = - \frac{\sharpe}{\sigma} \int_0^t
\Timefnct(s) \ds + (\frac{\sharpe^2}{2} - r)t, \) and define $F(t,w)$
by
\begin{equation}\label{eq:def_F}
  F(t,w) = e^{A(t)} \exp \left\{ - \frac{\sharpe}{\sigma} \int_1^{w}
    \frac{\di \xi}{\pits(t,\xi)} \right \}.
\end{equation}
By assumption, for $t>0$, $F(t,0)=\infty$ and $F(t,\infty)=0$.  For
each $t$, $F(t,w)$ is $\classC^{1,2}$ and decreasing in $w$, so we can
define its inverse $f=F^{-1}$ such that $f(t,F(t,w))=w$ and $F(t,f(t,z))=z$.

Finally set $H(t,c) = \int_1^{c} F(t, Y(t,b)) \di b$.  Note that we
have
\begin{eqnarray}
  \label{wFa}
  f_z(t,z) F_w(t,f(t,z)) & = & 1 ; \\
  \label{wFb}
  f_t(t,z) + f_z(t,z) F_t(t,f(t,z)) &=& 0 ; \\
  \label{wFc}
  f_z(t,z) F_{ww}(t,f(t,z)) + f_{zz}(t,z) F_w(t,f(t,z))^2 & = & 0 ,
\end{eqnarray}
and that $f$ is $\classC^{1,2}$.

\begin{theorem}
  \label{thm:Dual}
  For any $x>0$, the following two are equivalent:
  \begin{enumerate}
  \item $c(t,W^{x}_t)$ and $\pi(t,W^{x}_t)$ achieve a finite maximum
    in the problem
    \begin{equation}\label{eq:stocutility}
      \max_{C,\Pi \in \sA^u(x)} \E\left[ \int_0^\infty u(t,C_t) \dt \right],
    \end{equation}
    for a regular utility function $u$, as in Definition
    \ref{def:utility}, for which
    \begin{equation}\label{eq:ExistsLM}
      \exists \lambda { > 0} \textrm{ such that }
      x=\E\left[\int_0^\infty Z_t I(t,\lambda Z_t) \dt\right].
    \end{equation}
  \item $c(t,w),\pi(t,w)$ are a regular consumption/investment pair,
    as in Definition~\ref{def:regularpic}, $c(t,w),\pi(t,w)$ satisfy
    \eqref{eqn:black} on $(0,\infty)^2$ and
    \begin{equation}
      \label{eq:Budget} \E \left[ \int_0^\infty Z_t c(t,W^{x}_t) \dt \right] = x,
    \end{equation}
    and for some $0<x_0\leq x$, $\E[|H(t,c(t,W^{x_0}_t))|]<\infty$ for
    almost all $t \ge 0$ and $\int_0^\infty \E[ H(t,c(t,W^{x_0}_t)) -
    h(t) ]^+ \dt<\infty$, where $h(t)=\E[H(t,c(t,W^{x_0}_t))]$.
  \end{enumerate}
  Moreover, we then have $u_c(t,c)=H_c(t,c)$, $\sA^u(x)=\sA(x)$ and in
  (i) one may take $u(t,c)=H(t,c)-h(t)$.
\end{theorem}

\begin{remark}\label{rk:wealth_expression}
  In $(ii)$ it is equivalent to use $f(t, F(0,x)Z_t)$ in place of
  $W^x_t$ throughout. This condition may be easier to check.
\end{remark}

\begin{remark}\label{rk:det_stoch}
  It is interesting to observe the analogy with the deterministic
  setup considered in Section \ref{sec:deterministic}. There, given
  {an} agent's consumption, we recovered their utility function as
  $u_c(t,c)= \tilde{F}(t,y(t,c))$, where $y(t,c)$ was the inverse of
  consumption and $\tilde{F}(x)=\int_x^\infty D(s)\dd s$ was an
  arbitrary absolutely continuous decreasing non-negative function. In
  Theorem \ref{thm:Dual} above, we recover the utility function in the
  same form $u_c(t,c)=F(t,Y(t,c))$ but now $F$ is uniquely specified
  in terms of {the} agent's investment strategy coupled with the
  discounting term $A(t)$ read {off} from Black's equation
  \eqref{eqn:black}.
\end{remark}

\begin{remark}
There are close parallels between different conditions in $(i)$ and
  $(ii)$:
  \begin{itemize}
  \item The fundamental point of the theorem is the equivalence
    between \eqref{eqn:black} and optimality of $c,\pi$ for the
    problem \eqref{eq:stocutility}. If \eqref{eqn:black} fails,
    $c,\pi$ may still be optimal but for a more general problem of the
    type \eqref{eq:vfdefnmod}.
  \item The integrability conditions on $\pi(t,w)$: $\int_{0}^1 \di
    \xi / \pi(t,\xi)= \infty = \int_1^{\infty} \di \xi / \pi(t,\xi)$
    correspond to the Inada condition on $u$.
  \item Equations \eqref{eq:ExistsLM} and \eqref{eq:Budget} are
    essentially the same and encode the budget constraint.  We show
    below that if $\E [ \int_0^\infty Z_t c(t,W^{x}_t)\dt ]>x$ then
    $(C,\Pi)$ is not admissible. Conversely, if $\E [ \int_0^\infty
    Z_t c(t,W^{x}_t)\dt ]=x - \delta$ for $\delta>0$ then $c,\pi$ is
    typically not optimal. Indeed, if $\tilde{c}(t,w) = c(t,w) +
    \delta e^{-s}/Z_s$ then (by Theorem III.9.4 of
    \cite{KaratzasShreve:98}) there exists a process $\tilde{\Pi}$
    such that $\tilde{c},\tilde{\Pi}$ is admissible, and achieves a
    strictly higher expected utility of consumption over time in
    \eqref{eq:stocutility}.
  \end{itemize} In general {the} assumptions of Theorem \ref{thm:Dual}
  may be non-trivial to verify. However we provide a wide class of
  examples where they hold, see Lemma \ref{lem:globlip} below. In
  Section~\ref{sec:bdd}, we shall also discuss some ways in which the
  conditions of the theorem may be relaxed.
\end{remark}

\begin{remark}
  \label{rk:BlackHH}
  The focus in \citet{CoxLeland:00} and \citet{HeHuang:94} is on a
  problem in which \eqref{eq:stocutility} is replaced by maximisation
  of expected utility of consumption and terminal wealth over a finite
  horizon $[0,T]$. These papers give analogues of Theorem
  \ref{thm:Dual} above in this setting (\cite[Proposition
  3]{CoxLeland:00}, \cite[Theorems 1\&3]{HeHuang:94}).
  \citet{CoxLeland:00} develop discrete-time arguments, and pass to
  the limit without full justification to deduce the continuous-time
  result.  \citet{HeHuang:94} work directly in continuous time, and
  give a rigorous derivation of the results.

In many respects the infinite horizon problem is more natural than the
finite horizon version, but it introduces new difficulties related in
particular to the budget constraint
\eqref{eq:ExistsLM}--\eqref{eq:Budget} and finiteness of the value
function \eqref{eq:stocutility}.  \cite{HeHuang:94} comment that their
analysis could extend to an infinite horizon with the additional
condition $\E[Z_t W_t]\to 0$ as $t\to\infty$ which, under all their
assumptions, implies our budget constraint \eqref{eq:Budget}. However
this does not seem to be so immediate due to the important
integrability restrictions on $\cts$ in \citet{HeHuang:94}. Further,
the well-posedness and finiteness of the expected utility is not
discussed.

In addition to the finite horizon/infinite horizon distinction, we
believe our approach has the advantage of a mathematically rigorous
and unified \emph{if and only if} statement with a straightforward
proof which should be appealing to a contemporary reader. Our proof of
Theorem \ref{thm:Dual} is based on the dual approach. In contrast
\cite[Theorem 1]{HeHuang:94} use a primal approach to prove results
for the forward problem ((i) implies (ii) in our theorem), and a dual
approach for the inverse problem \cite[Theorem 3]{HeHuang:94}. The
mixing of primal and dual techniques can easily lead to
incompatibilities between sets of assumptions, and for this reason
\citet{HeHuang:94} do not have an if and only if statement. For
example, the assumptions of Theorem 3 in \cite{HeHuang:94} are easily
satisfied by consumption and investment strategies which are
proportional to wealth and which result from CRRA utility, see Example
\ref{ex:CRRAstoch} below. However, taking parameters which correspond
to risk aversion $\gamma>1$ means that the value function behaves as
$\frac{1}{1-\gamma}x^{1-\gamma}$ and does not satisfy the polynomial
growth restriction required for their Theorem 1.  The authors seemed
to have been aware of such instances, see Footnote~20 therein.

  We note also that we are able to make less restrictive assumptions
  than previous works. In particular, both \cite{CoxLeland:00} and
  \cite{HeHuang:94} assumed stronger growth and differentiability
properties on $\cts$ and $\pits$. Further, these properties were
imposed as standing
  assumptions for their theorems whereas in one direction we deduce
these properties from the regularity of $u$.
Moreover, our setup allows us to obtain a general class of actions for which we can
  verify the assumptions, see Lemma \ref{lem:globlip} below, which
  includes interesting examples. This seems very difficult in the set-up
of \citet{HeHuang:94}. 
  %

Nevertheless, we stress that \citet{HeHuang:94} remains a very
impressive paper, with many contributions which are beyond the scope
of this work. In particular, they considered a more general setup than
we do in that they allowed the stock price $P_t$ to be a generic
diffusion (local volatility) process and $\cts$ and $\pits$ to depend
on the state (i.e. $P_t$) as well.
\end{remark}
\begin{proof}[Proof of Theorem~\ref{thm:Dual}]
  We first show that $(ii)\Rightarrow (i)$. \\
  We take $u(t,c)=H(t,c)-h(t)$ which is strictly increasing and
  strictly concave. We have $u_c(t,0)=F(t,Y(t,0))=F(t,0)=\infty$,
  $u_c(t,\infty) =F(t,Y(t,\infty)) =F(t,\infty)=0$ and
  $u_{cc}(t,l)=F_w(t,Y(t,l))/c_w(t,Y(t,l))$ is continuous on $(0,
  \infty)$. In addition, $I(t,z) = c(t,f(t,z))$ and, as observed
  above, $f$ is well defined and $\classC^{1,2}$ on $(0,\infty)^2$.
  In consequence, $I_z(t,z)$ is locally H\"older continuous and $u$ is
  a regular utility function of Definition \ref{def:utility}.

  Let $\lambda=\lambda(x)=F(0,x)$ and set $W_t = f(t,\lambda(x)
  Z_{t})$, so that $W_0 = f(0,\lambda(x))=x=W^{x,\pi,c}_0$. We now
  show that $W_t = W^{x,\pi,c}_t$.

  Note that by construction we have $u_c(s, c(s,W_s)) = F(s, W_s)=
  \lambda(x)Z_s$. By It\^o's Lemma
  \begin{equation*}
    \begin{split}
      \di W_t = & \lambda(x) {f_z(t, \lambda(x) Z_t)} \di Z_t + f_t(t,
      \lambda(x) Z_t) \dt +
      \frac{\lambda(x)^2}{2} f_{zz}(t, \lambda Z_t)d\langle Z\rangle_t \\
      = & - \sharpe \lambda(x) Z_t f_z(t, \lambda(x) Z_t) \di B_t \\ & {} +
      \left( f_t(t, \lambda(x) Z_t) - r \lambda(x) Z_t f_z(t,
        \lambda(x) Z_t) + \frac{\sharpe^2 \lambda(x)^2}{2} Z_t^2
        f_{zz}(t, \lambda(x) Z_t) \right) \dt.
    \end{split}
  \end{equation*}
  Then $W_t$ is a strong solution to \eqref{eq:wealth_dynamics}
  provided that for $w=f(t,z)$,
  \[ - \sharpe z f_z(t,z) = \sigma \pi(t,w) \] and
  \[ \sharpe \sigma \pi(t,w) + rw - c(t,w) = f_t(t,z) - rz f_z(t,z) +
  \frac{1}{2} \sharpe^2 z^2 f_{zz}(t,z). \] For the first of these,
  using $z=F(t,w)$ and the definition of $F$ and (\ref{wFa}), we have
  \[ - \sharpe z f_z(t,z) = - \sharpe \frac{F(t,w)}{F_w(t,w)} =
  \sigma\pi(t,w).
  \]
  For the second, using also (\ref{wFb}) and (\ref{wFc}),
  \begin{eqnarray}
    \lefteqn{f_t(t,z) - rz f_z(t,z) + \frac{1}{2} \sharpe^2 z^2 f_{zz}(t,z)}
    \nonumber \\
    & = & - {f_z(t,z)} \left( F_t(t,w) + rF(t,w) + \frac{\sharpe^2}{2}
      \frac{F(t,w)^2
        F_{ww}(t,w)}{F_w(t,w)^2} \right) \nonumber\\
    & = & \frac{-F(t,w)}{F_w(t,w)} \left( A'(t) + \frac{\sharpe}{\sigma}
      \int_1^w \frac{\pi_t(t,\xi)}{(\pi(t,\xi))^2} d \xi + r + \frac{1}{2}
      \left[
        \sharpe^2 + \sigma \sharpe \pi_w(t,w) \right] \right)
    \label{eq:Ftused}\\
    & = & \frac{\sigma \pi(t,w)}{\sharpe} \left( \sharpe^2 + \frac{\sharpe
        r
        w}{\sigma \pi(t,w)} -
      \frac{\sharpe c(t,w)}{\sigma \pi(t,w)} \right) \nonumber\\
    & = & rw + \sharpe \sigma \pi(t,w) - c(t,w).
  \end{eqnarray}
  We thus conclude that $W_t^x$ and $W_t$ are strong solutions to the
  same SDE, \eqref{eq:wealth_dynamics}. By
  \citet[Theorem~5.2.5]{KaratzasShreve:91}, we therefore have $W^{x}_t
  = W_t = f(t,\lambda(x)Z_{t})$ for all $t \ge 0$ a.s.\ with
  $\lambda(x)=F(0,x)$.

  For the rest of the proof, with slight abuse of notational
  conventions, let us write $c_t^{x}:=c(t,W^x_t)$.  It follows that
  $c^x_t=c(t,f(t,F(0,x)Z_{t}))$ a.s.\ and in particular $c_s^{x}\leq
  c_s^{y}$ for $0<x<y$. Further, since
  $u_c(t,c(t,f(t,z)))=F(t,f(t,z))=z$, we have
  $u_c(t, c_t^x) = \lambda(x) Z_{t}$, so that $c^x_t = I(t, \lambda(x)
  Z_t)$ and
  \begin{equation}
    \label{fo2}
    \tilde{u}(t,\lambda(x)Z_{t})=u(t,c_t^x)-\lambda(x)Z_{t}c_t^x,
  \end{equation}
  where $\tilde{u}$ is the convex dual of $u$. It follows that
  \eqref{eq:ExistsLM} is simply \eqref{eq:Budget}.

  By the assumption $\E[|u(t,c(t,W^{x_0}_t))|] \leq
  \E[|H(t,{c}(t,W^{x_0}_t))|] + |h(t)| < \infty$ and
  \[ \E[u(t,c(t,W^{x_0}_t))] = \E[H(t,c(t,W^{x_0}_t))] - h(t) = 0. \]
  Using the hypothesis $\E[\int_{0}^{\infty} [u(t,c(t,W^{x_0}_t))]^+
  \dt] < \infty$ we obtain
  \[\E[\int_{0}^{\infty} u(t,c(t,W^{x_0}_t)) \dt] = 0.
  \]
  For $x>x_0$ we write
  \begin{equation*}
    u(t,c_t^{x})\leq \tilde{u}(t,\lambda(x_0)Z_t)+\lambda(x_0)Z_tc_t^{x}=u(t,c_t^{x_0})+\lambda(x_0)(c_t^{x}-c_t^{x_0})Z_t
  \end{equation*}
  and hence
  \begin{equation*}
    \E \int_0^\infty u(t,c_t^{x})^+ \dt\leq \E \int_0^\infty u(t,c_t^{x_0})^+ \dt + \lambda(x_0)\E \int_0^\infty Z_t c^{x}_t\dt<\infty.
  \end{equation*}
  Hence $\E[\int_{0}^{\infty} u(t,c^{x}_t) \dt]$ is well defined and
  non-negative.

  Consider now arbitrary $C,\Pi \in \sA(x)$. From
  \eqref{eq:ZWdynamics} we have
  \[ 0 \leq W_tZ_t = x + \int_{0}^t Z_s( \sigma \Pi_s - \sharpe W_s)
  \di B_s - \int_0^t Z_s C_s \ds . \] It follows that
  \[ 0 \leq \int_0^t Z_s C_s\ds  \leq x + \int_{0}^t Z_s(\sigma \Pi_s -
  \sharpe W_s) \di B_s . \] In particular $\int_{0}^t Z_s( \Pi_s -
  \sharpe W_s) \di B_s \geq - x$ is bounded below and hence is a
  supermartingale. We also conclude that for each $t$, $\E[ \int_0^t
  Z_s C_s\ds  ] \leq x$ and hence
  \begin{equation}\label{eq:budget_bound}
    \E\left[ \int_0^\infty Z_s C_s \ds \right] \leq  x.
  \end{equation}
  It follows that, with $\lambda=\lambda(x)$,
  \begin{eqnarray*}
    \E\left[ \int_{0}^{\infty} u(s,C_s)^+ \ds \right] & \leq &
    x \lambda  + \E\left[ \int_{0}^{\infty} (u(s,C_s) - \lambda Z_s C_s)^+ \ds \right] \\
    & \leq &  x \lambda + \E\left[\int_{0}^{\infty} \tilde{u}(s,\lambda Z_s)^+ \ds\right]\\
    & = &  x \lambda + \E\left[\int_{0}^{\infty} (u(s,c_s^{x}) - \lambda Z_s c_s^{x})^+ \ds\right] \\
    &\leq &  x \lambda + \E\left[\int_{0}^{\infty} u(s,c_s^{x})^+ \ds\right] < \infty,
  \end{eqnarray*}
  where we used \eqref{fo2}. In consequence, $\sA(x)=\sA^u(x)$.  Once
  we know the expectations exist we proceed with a standard argument:
  \begin{equation}
    \begin{split}
      \E\left[ \int_0^\infty u(s,C_s) \ds \right]
      \leq& \ \lambda x + \E\left[ \int_0^\infty u(s,C_s) - \lambda Z_s C_s  \ds \right]  \\
      \leq & \ \lambda x + \E\left[ \int_0^\infty \tilde{u}(s,\lambda
        Z_s) \ds \right]\ .
      \label{firstorder}
    \end{split}
  \end{equation}
  Further, from (\ref{eq:Budget}) and \eqref{fo2} it is immediate that
  there is equality throughout \eqref{firstorder} for $C_s=c_s^{x}$
  and $\Pi_s=\pi(s,W_s^{x})$ which shows that these are optimal.
  \smallskip\\
  We come now to the other implication: $(i)\Rightarrow(ii)$.\\
  Take $\lambda$ as in \eqref{eq:ExistsLM} and let $C_s:=I(s,\lambda
  Z_s)$. In particular, $\tilde{u}(s,\lambda Z_s)=u(s,C_s)-\lambda
  Z_s C_s\geq u(s,c^x_s)-\lambda Z_s c^x_s$ and hence
  $$u(s,c^x_s)^-\geq \left(\lambda Z_s(c^x_s-C_s)+u(s,C_s)\right)^-\geq u(s,C_s)^--\lambda Z_s(c_s^x-C_s)^+.$$
  Rearranging and integrating we have
  $$\E\left[ \int_0^\infty u(s,C_s)^- \ds \right]\leq \E\left[ \int_0^\infty u(s,c^{x}_s)^- \ds \right] + \lambda\E\left[ \int_0^\infty Z_s(c_s^x+C_s) \ds \right] < \infty,$$
  where we used \eqref{eq:budget_bound} and the fact that $c^x_s$
  induces a finite maximum in \eqref{eq:stocutility}.
    As observed earlier (cf.~Theorem III.9.4 in
  \cite{KaratzasShreve:98}), there exists $(\Pi_s)$ such that
  $(C_s,\Pi_s)\in \sA(x)$ and the above then shows that
  $(C_s,\Pi_s)\in \sA^u(x)$.
  Proceeding as in \eqref{firstorder}, we obtain
  $$
  \E\left[ \int_0^\infty u(s,c^{x}_s) \ds \right] \leq \lambda x +
  \E\left[ \int_0^\infty \tilde{u}(s,\lambda Z_s) \ds \right] =
  \E\left[ \int_0^\infty u(s,C_s) \ds \right].
  $$
  It follows we have to have equality in the above {equation} which is
  true if and only if $u(s,c^x_s)-\lambda Z_s
  c^x_s=\tilde{u}(s,\lambda Z_s)$ $\ds\times \di \Pr$-a.e., which in
  turn is true if and only if $c^{x}_s=C_s$ {almost surely}. In
  consequence, \eqref{eq:Budget} is simply \eqref{eq:ExistsLM}.

  Using similar arguments to the ones which led to
  \eqref{eq:budget_bound} above we see that $W_tZ_t\to W_\infty
  Z_\infty$ a.s.\ as $t\to\infty$. Further
  $$0\leq \E W_\infty Z_\infty\leq x-\E\int_0^\infty Z_s C_s\ds=0$$
  and hence $W_\infty Z_\infty\equiv 0$. In addition, by considering a
  localising sequence of stopping times $\tau_N$, from
  \eqref{eq:ZWdynamics} we get:
  \begin{equation*}
    x = \E\left[ W_{\tau_N} Z_{\tau_N}\right] + \E \left[
      \int_0^{\tau_N} C_s Z_s \ds \right].
  \end{equation*}
  But from \eqref{eq:ExistsLM} $x = \E\int_0^\infty Z_s C_s\ds$, so
  $\E\left[ W_{\tau_N} Z_{\tau_N}\right] \to 0$ as $N \to \infty$.
  Moreover,
  \begin{equation*}
    W_t Z_t = \E \left[\int_t^{\tau_N} C_s Z_s \ds \Big| \Fc_t\right]
    + \E \left[ W_{\tau_N} Z_{\tau_N}\big| \Fc_t\right],
  \end{equation*}
  and the final term is almost surely non-negative. So
  \begin{equation*}
    W_t Z_t - \E \left[\int_t^{\infty} C_s Z_s \ds \Big| \Fc_t\right]
    \ge 0
  \end{equation*}
  almost surely, and taking expectations, we see that this has
  expected value $\lim_{N \to \infty} \E\left[W_{\tau_N}
    Z_{\tau_N}\right] =0$. Hence
  \begin{equation*}
    W_t = \frac{1}{Z_t} \E \left[\int_t^{\infty} C_s Z_s \ds \Big|
      \Fc_t\right] = \frac{1}{Z_t} \E \left[ \int_t^\infty
        I(s,\lambda Z_s) Z_s \ds \Big| \Fc_t\right].
  \end{equation*}


    Define $g(t,z)$ to be the solution to the PDE
    \begin{equation*}
      g_t + z (\theta^2-r) g_z + \half z^2 \theta^2 g_{zz} - rg =
      -I(t,\lambda z)
    \end{equation*}
    with initial condition $g(0,z) = \E \left[\int_0^\infty
      I(s,\lambda Z_s) Z_s \ds \big| Z_0 = z \right]$. It follows that
    $g\in \classC^{1,2}$, and in fact (\eg{}
    \citet[Theorem~3.5.10]{Friedman:64}) since $I_z$ is locally
    H\"older continuous, that $g_{tz}$ exists and is also locally
    H\"older continuous. Applying It\^o's Lemma, and using a similar
    localisation and convergence argument as above, we deduce that
    \begin{equation*}
      g(t,Z_t) = \frac{1}{Z_t} \E \left[ \int_t^\infty  I(s,\lambda
        Z_s) Z_s \ds | \Fc_t\right].
    \end{equation*}
    We conclude that $W_t = g(t,Z_t)$ and since $I$ is strictly
    decreasing on $(0,\infty)$ in $Z_t$, and noting that $Z_s/Z_t$ is
    independent of $Z_t$, then $g(t, \cdot)$ is also strictly
    decreasing.  Additionally, $g$ is strictly positive and both
    $g(t,z)$ and $z g_z(t,z)$ tend to zero as $z$ tends to infinity,
    and $g(t,z)$ tends to infinity as $x$ tends to 0. We have $c(t,w)
    = I(t,\lambda g^{-1}(t,w))$, and we conclude that $c(t,0)=0$,
    $c(t,\cdot)$ is strictly increasing, unbounded and $c_w(t,w)$ is
    locally H\"older continuous on $(0,\infty)^2$. Finally, we deduce
    that $g = Y \circ I$.

    Using It\^o's Lemma and equating $\di W_t$ with the wealth
    dynamics in \eqref{eq:wealth_dynamics} we obtain
    \eqref{pidef}--\eqref{Cdef}, with $g$ instead of $f$, which hold
    for all $t,z>0$. It follows that $\pi(t,0)=0$, and $\pi(t,w)>0$
    for $w>0$.  From \eqref{pidef}, and (as noted above) since
    $g_{tz}$ exists, we get the required differentiability properties
    of $\pi$. We then proceed as in
    \eqref{w''eqn}--\eqref{picfundeqn1}, to conclude that
    \eqref{eqn:black} holds.

    This means $F$ in \eqref{eq:def_F} is well defined and we may
    consider $\tilde{W}_t=f(t,F(0,x) Z_t)$, with $f=F^{-1}$ as
    above. Proceeding as in the first part of the proof it follows
    that $\tilde{W}_t$ is a strong solution to the SDE
    \eqref{eq:wealth_dynamics} considered for $0\leq t<\tau$, where
    $$\tau=\inf\{s: Z_s\notin
    (F(s,\infty)/F(0,x),F(s,0)/F(0,x))\}=\inf\{s: \tilde{W}_s \in
    \{0,\infty\}\}.$$ Unicity of strong solution to an SDE, as invoked
    above, holds also when we consider the SDE not on $t\in
    (0,\infty)$ but on $[0,\tau)$, and we conclude that
    $\tilde{W}_t=W_t^{x}=g(t,\lambda Z_t)$.  However since we know
    that $0<W_t^{x}<\infty$ a.s.\ it follows that $\tau=\infty$; i.e.\
    $F(t,0)=\infty$ and $F(t,\infty)=0$.  Finally, from
    $c^x_t=I(t,\lambda Z_t)$ and $W^x_t=\tilde{W}_t$, it also follows
    that $u_c(t,c)=H_c(t,c)$ so that $u(t,c)=H(t,c)-\zeta(t)$, for
    some function $\zeta$.  As $c_t^x$ achieves a finite maximum in
    \eqref{eq:stocutility} {it} follows that
    {$\E[|H(t,c(t,W^{x}_t))|]<\infty$} for a.e.\ $t\geq 0$ and
    Fubini's theorem yields
  $\int_0^\infty(h(t)-\zeta(t))\dt$ is well defined and finite {when we
    take $h(t) = \E[H(t,c(t,W^{x}_t))]$}. In consequence,
    \begin{equation}
\begin{split}
\E\int_0^\infty [ H(t,c(t,W^{{{x}}}_t)) - h(t) ]^+
  \dt<\E\int_0^\infty [ & H(t,c(t,W^{{{x}}}_t)) - \zeta(t) ]^+ \dt \\
  & + \int_0^\infty (\zeta(t)-h(t))^+\dt<\infty .
\end{split}
\end{equation}
  Hence {\it (ii)} holds when we take $x = x_0$. It follows from the first part of
  the proof that we may take $u(t,c)=H(t,c)-h(t)$.
\end{proof}

\subsection{Example with $c, \pi$ linear in wealth}

\begin{example}\label{ex:CRRAstoch}
  Suppose $c(t,w) = \kappa w$ and $\pi(t,w)= \phi w$ for $\kappa, \phi
  >0$ with $\phi \neq \sharpe/\sigma$.  Then Black's equation
  (\ref{eqn:black}) is satisfied, $Y(t,b)=b/\kappa$, $\Timefnct(t)
  \equiv \Timefnct = (\kappa - r)/ \phi + \sigma^2 \phi/2$ and $A(t) =
  \xi t$ where $\xi = - \sharpe \Timefnct/\sigma + \sharpe^2/2 - r$.

  Let $\ra=\sharpe/\phi \sigma$. We have $F(t,w) = e^{\xi t} w^{-\ra}$
  and in particular $F(t,0)=\infty$, $F(t,\infty)=0$. It follows that
  $\lambda(x)=x^{-\ra}$ and $f(t,z) = z^{-1/\ra} e^{ (\xi / \ra)t}$,
  which is $\classC^{1,2}$ differentiable.

  Further, $W^{x}_t = f(t, \lambda(x)Z_t) = xe^{ \phi \sigma B_t +
    (\sigma \phi \sharpe + r - \kappa - \sigma^2 \phi^2/2)t}$ and a
  direct computation yields
  \begin{equation}\label{eq:crraex_compute}
    \me^{\xi t}\E[ (W_t^{x})^{1 - \ra}] = x^{1-\ra}\me^{\xi t/\ra}\E[Z^{1-1/\ra}_t]=x^{1-\ra}\me^{-\kappa t}.
  \end{equation}
  It follows that
  \[ \E\left[ \int_0^\infty Z_t c(t,W^{x}_t) \dt\right] = \kappa x
  \E\left[ \int_0^\infty (Z_t)^{1 - 1/\ra} \me^ {\xi t/\ra} \dt \right]
  = \kappa x \int_0^\infty \me^{-\kappa t} \dt = x.
  \]
  Also $H(t,c)= \frac{1}{1-\ra}\me^{\xi t}\kappa^{\ra} [c^{1 -\ra} -
  1]$ so that, taking $x_0=1$,
  \[ h(t) = \E[H(t,c(t,W^{x_0}_t))] =
  \frac{1}{1-\ra}\left({\kappa}\me^{-\kappa t} - \me^{\xi t}
    \kappa^\ra\right),
  \]
  and
  \begin{equation}
    \label{udefn}
    \begin{split}
      u(t,c) & = \frac{1}{1-\ra}\left(\kappa^\ra\me^{\xi t}c^{1-\ra}-\kappa \me^{-\kappa t}\right)\\
      & = \frac{\kappa^\ra\me^{\xi t}}{1-\ra}\left(c^{1-\ra}-(\kappa
        \me^{-\zeta t})^{1-\ra}\right),
    \end{split}
  \end{equation}
  where $\zeta= (\xi + \kappa)/(1-\ra)=\kappa - r
  -\frac{\theta^2}{2\ra}$. Then, using \eqref{eq:crraex_compute},
  $\E[u(t,c(t,W^{x_0}_t))^+] < D \me^{-\kappa t}$ for some constant
  $D$, and hence $\int_0^\infty \E [H(t,C_t)-h(t)]^+ \dt < \infty$ for
  the optimal policy.

  In the above we could take any $x_0>0$. So, in conclusion, for any
  initial capital $x>0$, $\pi,c$ solve the optimal consumption problem
  for admissible strategies for $u$ as given in \eqref{udefn}.  We
  note that the choice of consumption and investment which are linear
  in wealth and time-homogeneous necessarily implies an exponential
  discounting of utility from a given wealth.

\end{example}

\section{Consequences and Extensions of the Main Result} 
\label{sec:consequences}

\subsection{Model uncertainty}\label{sec:model_uncert}
In our analysis so far we assumed agents believe that the price
process follows the Black-Scholes model \eqref{eq:price_dynamics} with
given parameters $\sharpe,\sigma$. We then asked whether their
observed actions are optimal for \eqref{eq:stocutility} for
\emph{some} utility function $u$.  Suppose however that we do
\emph{not} know agents' beliefs about model parameters.  We may then
ask more generally whether agents' actions are optimal for \emph{some}
utility function $u$ and \emph{some} price dynamics?\footnote{We are
  grateful to Masaaki Fukasawa for suggesting this
  question. See also \citet{CuocoZapatero:00} for related results,
  although with an emphasis on equilibrium constraints.}

More precisely, within the realm of Theorem \ref{thm:Dual}, we can ask
the following: are the observed actions optimal for
\eqref{eq:stocutility} for \emph{some} $u$ and \emph{some}
$\sharpe,\sigma$? Assume we are not in the special case when
$\pits_w(t,w)=\phi(t)$. Then $\pits$ and $\cts$ must solve Black's PDE
\eqref{eqn:black} and clearly there can be at most one value of
$\sigma$ for which \eqref{eqn:black} is satisfied. Put differently, if
we find that agents are optimising expected utility of consumption
then we also recover uniquely their belief about market's
volatility. In contrast we do not recover their belief about the
Sharpe ratio $\sharpe$, which does not appear in \eqref{eqn:black}.
Indeed, as we argue below, if $c,\pi$ are consistent with utility
maximisation for a model with Sharpe ratio $\sharpe$, then we
expect that they are also consistent with utility maximisation for a
model with a different Sharpe ratio $\hat{\sharpe}$, but for a different
utility.

Consider $\hat \Pr$, $\hat Z_t$ defined via
$$\frac{\di \hat \Pr}{\di \Pr}\Big|_{\F_t}=\exp\left((\hat\sharpe-\sharpe)B_t-\frac{(\hat\sharpe - \sharpe)^2}{2}t\right)=\frac{Z_t}{\hat Z_t}$$
so that
$$\frac{\di P_t}{P_t}=\sigma(\di \hat B_t + \hat\sharpe \dt)+ r \dt,$$
for a $\hat\Pr$--Brownian motion $\hat B_t=
B_t-(\hat\sharpe-\sharpe)t$. Observe further that $Z_t=\hat
Z_t^{\sharpe/\hat\sharpe}\mathrm{e}^{\mu t}$, with
$\mu=\frac{\sharpe}{2}(\sharpe-\hat\sharpe)+r(\frac{\sharpe}{\hat\sharpe}-1)$.

Suppose agents' actions $\pits,\cts$ are given and the equivalent
conditions in Theorem \ref{thm:Dual} hold. Define a new utility
function $\hat u(t,c)$ via the inverse of $\hat u_c$: $\hat
I(t,z):=I(t,z^{\sharpe/\hat\sharpe}\mathrm{e}^{\mu t})$. It follows
that the budget equation \eqref{eq:ExistsLM} holds for $\hat\lambda =
\lambda^{\hat\sharpe/\sharpe}$. Indeed, by definition,
$$\hat\E \left[\int_0^\infty \hat Z_t \hat I(t,\hat \lambda \hat Z_t)\dt\right]=
\hat\E \left[\int_0^\infty \hat Z_t I(t, \lambda Z_t)\dt\right]=
\E\left[\int_0^\infty Z_t I(t,\lambda Z_t)\dt\right]=x.$$ Classical
duality arguments (see the proof of the implication $(i)\Rightarrow
(ii)$ in Theorem \ref{thm:Dual} above), yield that the optimal
consumption policy in the problem \eqref{eq:stocutility} for $\hat
u$ under $\hat \Pr$ is given as
$$C_t=\hat I(t,\hat \lambda \hat Z_t)=I(t,\lambda Z_t)=c(t,W^x_t).$$
Provided \eqref{eq:stocutility} for $\hat u$ under $\hat \Pr$
has a finite value, we conclude that the agents' chosen actions are
optimal for $u$ under $\Pr$ and for $\hat u$ under $\hat \Pr$.

We will not persue this finiteness issue here. However,
Lemma \ref{lem:globlip} below
gives sufficient conditions under which the problem value is finite, and
then we
conclude that the same consumption/investment pair is consistent with
a family of utility functions, each member of which corresponds to a
different model and Sharpe ratio.

\subsection{Risk aversion}\label{sec:riskaversion}
We now return to our original setting where $\sigma,\sharpe$ are known
and fixed and we consider questions similar to those which arose in
Theorem~\ref{thm:ARAsign}, namely, we investigate what we can say
about an investor's risk profile from her actions. Recall the absolute
risk aversion $\rho(t,c)$ given in Definition \ref{def:ra}.
\begin{prop}\label{prop:riskaversion}
  An investor with investment and consumption strategies $\pits(t,w)$,
  $\cts(t,w)$ satisfying the assumptions of {\it (ii)} of Theorem
  \ref{thm:Dual}, where $\cts(t,w)$ is twice continuously
  differentiable in wealth, has decreasing absolute risk aversion
  (DARA) if and only if
  \begin{equation}\label{eq:DARA_stoch}
    \frac{\pits_w(t,w)}{\pits(t,w)}\geq -\frac{\cts_{ww}(t,w)}{\cts_w(t,w)}.
  \end{equation}
  In particular, a sufficient condition for an investor to be DARA is
  convexity (in wealth) of her consumption and investment which is
  increasing in wealth.
\end{prop}
\begin{proof} From the final statement of Theorem \ref{thm:Dual}, the
  investor's utility function $u$ satisfies
  $u_c(t,c)=H_c(t,c)=F(t,Y(t,c))$ and from \eqref{eq:def_F} it follows
  that
  \begin{equation*}
    \rho(t,c) = \frac{\sharpe}{\sigma\pits(t,Y(t,c))}Y_c(t,c).
  \end{equation*}
  The absolute risk aversion is decreasing, 
  \ie{} $\rho_c(t,c) \le 0$ iff:
  \begin{align*}
    && 0 \ge & \pd{}{c} \left[
      \frac{Y_c(t,c) }{\pits(t,Y(t,c))}\right]\\
    & \iff & 0 \ge & \frac{Y_{cc}(t,c)}{\pits(t,Y(t,c))} -
    \frac{\left(Y_c(t,c)\right)^2\pits_w(t,Y(t,c))}{\pits(t,Y(t,c))^2}
    \\
    & \iff & 0 \ge & - \left. \frac{\cts_{ww}(t,w)}{\pits(t,w)
        \cts_w(t,w)^3} - \frac{\pits_w(t,w)}{\cts_w(t,w)^2
        \pits(t,w)^2}\right|_{w =
      Y(t,c)}\\
    &\iff & 0 \le & \left. \pits_w(t,w) + \pits(t,w)
      \frac{\cts_{ww}(t,w)}{\cts_w(t,w)}\right|_{w = Y(t,c)}.
  \end{align*}
Transforming the last inequality
  we arrive at the statement of the proposition.
\end{proof}
In a similar manner we derive a condition equivalent to relative risk
aversion. We omit the proof for the sake of brevity.
\begin{prop}
  An investor with investment and consumption strategies $\pits(t,w)$
  and $\cts(t,w)$ satisfying the assumptions of {\it (ii)} of Theorem
  \ref{thm:Dual}, where $\cts(t,w)$ is twice continuously
  differentiable in wealth, has decreasing relative risk aversion
  (DRRA) if and only if
  \begin{equation*}
    \left(\cts_w(t,w)\right)^2 - \cts(t,w) \cts_{ww}({t,}w) \le
    \frac{\cts_w(t,w) \cts(t,w) \pits_w(t,w)}{\pits(t,w)}
  \end{equation*}
  or equivalently iff
  \begin{equation*}
    \pd{}{w}\left[\log \frac{\cts(t,w)}{\cts_w(t,w)} \right] \leq
    \pd{}{w}\log \pits(t,w).
  \end{equation*}
\end{prop}

\subsection{Time-homogeneous investment and consumption}\label{sec:timehom}
We specialise now to the important special case of $\pits(t,w)$,
$\cts(t,w)$ which are independent of time.  Suppose first that
$\pits(t,w)=\pits(w)$ is independent of time. Equation
\eqref{eq:cpiPDE2} then simplifies to
\begin{equation*}
  \rts_w(t,w)=\frac{r}{\pits(w)}-\frac{r
    w}{\pits(w)^2}\pits_w(w)-\frac{\sigma^2}{2}
  \pits_{ww}(w)=\pd{}{w}\left(\frac{rw}{\pits(w)}
    -\frac{\sigma^2}{2}\pits_w(w)\right)
\end{equation*}
which yields
\begin{equation*}
  \rts(t,w)=\frac{rw}{\pits(w)}-\frac{\sigma^2}{2}\pits_w(w)+\Timefnct(t),
\end{equation*}
where $\Timefnct(t)$ is taken such that $\rts(t,w)>0$. In consequence
\begin{equation}\label{eq:timehom_c}
  \cts(t,w)=\rts(t,w)\pits(w)=rw-\frac{\sigma^2}{2}\pits(w)\pits_w(w)+\Timefnct(t)
  \pits(w)\ .
\end{equation}
In particular, if the agent invests a constant proportion of wealth in
the risky asset, i.e.\ $\pits(w)=\phi w$, $\phi>0$, then
\begin{equation*}
  \cts(t,w)=\left(r-\frac{\sigma^2}{2}\phi^2+\Timefnct(t)\phi\right)w=\rts(t,1) \phi w
\end{equation*}
is also linear in wealth. It is straightforward to see that for a
reasonable $\Timefnct(t)$ (e.g.\
continuous and bounded) agent's choices $\cts$ and $\pits$ verify the
assumptions of Theorem \ref{thm:Dual}. The case $\Timefnct(t)\equiv
\Timefnct$, a constant, was worked out explicitly in Example
\ref{ex:CRRAstoch} above.

We want to study in more detail the implications of representation
\eqref{eq:timehom_c} on the possible behaviour of admissible
investment/consumption strategies. Assume that $c=c(w)$ is also
time-homogenous, or equivalently that $\Timefnct(t)\equiv \Timefnct$
is a constant. Recall that we require $\pits(0)=\cts(0)=0$ and both
$\pits(w),\cts(w)$ are non-negative and $\cts(w)$ is
increasing. Consider an investment strategy given by $\pits(w)=\phi
w^\alpha$ with $\alpha>0$. Then \eqref{eq:timehom_c} gives
\begin{equation}\label{eq:opt_c_ex_power}
  \cts(w)=rw-\frac{\sigma^2\phi^2\alpha}{2}w^{2\alpha-1}+\Timefnct\phi w^\alpha.
\end{equation}
The condition $\cts(0)=0$ restricts us to $\alpha >1/2$. Considering
$\alpha \in (0.5,1)$ we see that the middle term in
\eqref{eq:opt_c_ex_power} dominates for small $w$ so that
$\cts_w(0+)=-\infty$ and $\cts(w)$ is negative for small values of
$w$. On the other hand, if $\alpha>1$ then the middle term dominates
for large values of $w$ and $\cts(w)$ becomes negative then. We
conclude that the only admissible value is the one studied above:
$\alpha=1$. This indicates that an admissible investment strategy has
to have linear behaviour near zero and infinity. For such actions we
are able to verify the assumptions in Theorem \ref{thm:Dual}.

\begin{lemma}\label{lem:globlip}
  Suppose $\cts(t,0)=0=\pits(t,0)$, $\pits(t,w)=\pits(w)$ is time
  homogeneous, $\cts(t,w)$ is continuous and $\cts, \pits$ are
  continuously differentiable in $w$ and $\cts_w$ is locally H\"older
  continuous on $(0,\infty)^2$. Further, $\cts$ and $\pits$ satisfy
  \eqref{eqn:black} and there exist strictly positive constants
  $\tilde\delta_1,\tilde\delta_2,\kappa_1,\kappa_2$ with
  \begin{equation}\label{eq:globlip}
    \begin{split}
      &\pits_w(w)\xrightarrow[w\to \infty]{} \tilde{\delta}_1,\quad \pits_w(w)\xrightarrow[w\to 0]{} \tilde{\delta}_2,\\
      & \delta_1 := \tilde{\delta}_1 \land \tilde{\delta}_2\leq
      \pits_w(w)\leq \tilde{\delta}_1\lor \tilde{\delta}_2 =:
      \delta_2,\textrm{ and } \kappa_1 \leq \cts_w(t,w)\leq
      \kappa_2,\, t\geq 0, w\geq 0.
    \end{split}
  \end{equation}
  Finally, assume either that $\sharpe/\sigma \leq \delta_1$, or
  $\delta_1 <\sharpe/\sigma \leq \delta_2$ and
  $(\sharpe(1-\delta_2/\delta_1) + \sigma \delta_2) > 0$. \footnote{We
    are grateful to Li Yu for noticing an error in an earlier version
    of this paper, in which we incorrectly stated a slightly different
    set of conditions.}

  Then $\cts,\pits$ is a regular investment/consumption pair
  (Definition \ref{def:regularpic}) and, for any $x>0$,
  $\E[|H(t,c(t,W^{x}_t))|]<\infty$ and $\int_0^\infty \E[
  H(t,c(t,W^{x}_t)) - h(t) ]^+ \dt<\infty$. Further, $F(t,0)=\infty$,
  $F(t,\infty)=0$, for all $t\geq 0$, and \eqref{eq:Budget} holds for
  any $x>0$. In consequence, (i) and (ii) in Theorem \ref{thm:Dual}
  hold true.
\end{lemma}

\begin{proof}
  It is immediate that $\int_0^1 d\xi/\pits(\xi) = \int_1^\infty
  d\xi/\pits(\xi)=\infty$ and hence $F(t,0)=\infty$ and
  $F(t,\infty)=0$. The other properties of Definition
  \ref{def:regularpic} follow equally easily. Fix $x>0$ for the rest
  of the proof.  Equation \eqref{eq:globlip} implies {global}
  Lipschitz behaviour of $\cts,\pits$ which guarantees the existence
  of a strong unique solution to \eqref{eq:wealth_dynamics}. Let $\Q$
  be the risk neutral measure under which $B^\sharpe_t:=B_t+\sharpe t$
  is a Brownian motion and put $\tilde{W}_t:=\me^{-rt} W_t$. We then
  have
  \begin{equation}
    \label{eq:tildeW}
    \tilde W_t =x + \sigma M_t - \int_0^t \me^{-rs}c(s,W_s)\ds,
  \end{equation}
  where $M_t:=\int_0^t \me^{-rs}\pi(W_s)\di B^\sharpe_s$ is a
  $\Q$-local martingale. In particular, $\tilde{W}_t$ is a
  non-negative super-martingale under $\Q$ and hence converges,
  $\Q$-a.s.\ as $t\to\infty$. It follows that $M_t$ also
  converges $\Q$-a.s.\ which is equivalent to $\langle
  M\rangle_t$ converging (cf.\ \cite[Proposition
  V.1.8]{RevuzYor:01}). However
  $$\int_0^t \delta_1^2 \tilde{W}_s^2 \ds\ \leq\  \langle
  M\rangle_t=\int_0^t \me^{-2rs}\pits(W_s)^2\ds\ \leq\ \int_0^t
  \delta_2^2 \tilde{W}_s^2 \ds$$ and it follows that $\tilde{W}_t\to
  0$ $\Q$-a.s. Finally, from classical estimates (e.g.\ \cite[Theorem
  5.2.3]{Friedman:75}), we have that $\E^\Q[(\tilde W_t)^m]<\infty$
  for all $m\geq 1$. It follows that $\E^\Q[\langle
  M\rangle_t]<\infty$ and hence $M_t$ is a $\Q$-martingale with $\E^\Q
  M_t=0$. In particular
  $$\E\int_0^\infty Z_s\cts(s,W_s)\ds=\lim_{t\to\infty}\E^\Q\int_0^t
  \me^{-rs}\cts(s,W_s)\ds=x - \lim_{t\to\infty}\E^\Q[\tilde W_t].$$ To
  show \eqref{eq:Budget}, it remains to argue that
  $\lim_{t\to\infty}\E^\Q[\tilde W_t]=0$. It follows from the above
  representation that $\E^\Q[\tilde W_t]$ is decreasing in $t$. By
  \eqref{eq:tildeW}, and $\E^\Q M_t = 0$, we have:
  \begin{equation*}
    \E^{\Q} \tilde{W}_t = x - \E^{\Q}\int_0^t \me^{-rs}
    c(s,\me^{rs}\tilde{W}_s)\ds.
  \end{equation*}
  Using the fact that $c(s,w) \ge \kappa_1 w$, and applying Fubini's
  theorem, we get:
  \begin{equation*}
    \E^{\Q} \tilde{W}_t \le x - \kappa_1 \int_0^t \E^{\Q} \tilde{W}_s \, \ds.
  \end{equation*}
  The desired conclusion follows immediately. It remains to show the
  integrability properties of $H$. We will show the stronger fact that
  $\E |H(t,c(t,W_t))| < \infty$ for all $t$, and also that $\E
  \int_0^\infty|H(t,c(t,W_t))|\, \dt < \infty$. From
  \eqref{eq:globlip} we get instantly that
  \begin{equation}
    \begin{split}
      &w^{-\frac{\sharpe}{\sigma\delta_2}}\leq \mathrm{e}^{-A(t)}F(t,w)\leq w^{-\frac{\sharpe}{\sigma\delta_1}},\quad 0\leq w<1\\
      &w^{-\frac{\sharpe}{\sigma\delta_1}}\leq
      \mathrm{e}^{-A(t)}F(t,w)\leq
      w^{-\frac{\sharpe}{\sigma\delta_2}},\quad w\geq 1
    \end{split}
  \end{equation}
  from which it follows that
  \begin{equation}\label{eq:f_estimates}
    \begin{split}
      &\mathrm{e}^{A(t)\frac{\sigma\delta_1}{\sharpe}}z^{-\frac{\sigma\delta_1}{\sharpe}}\leq
      f(t,z)\leq
      \mathrm{e}^{A(t)\frac{\sigma\delta_2}{\sharpe}}z^{-\frac{\sigma\delta_2}{\sharpe}},\quad 0\leq z<\mathrm{e}^{A(t)}\\
      &\mathrm{e}^{A(t)\frac{\sigma\delta_2}{\sharpe}}z^{-\frac{\sigma\delta_2}{\sharpe}}\leq
      f(t,z)\leq
      \mathrm{e}^{A(t)\frac{\sigma\delta_1}{\sharpe}}z^{-\frac{\sigma\delta_1}{\sharpe}},\quad
      z\geq \mathrm{e}^{A(t)}\ .
    \end{split}
  \end{equation}
  The integrability properties we need to establish are invariant
  under a shift of $H$ by a constant so we are free to redefine $H$ as
    $$H(t,z):=\int_{c(t,1)}^z F(t,Y(t,b))\di b, \textrm{ so that } H(t,c(t,w))=\int_1^w F(t,s)c_w(t,s)\ds.$$

    In consequence
    \begin{equation}\label{eq:H+}
      H(t,c(t,w))^+ = \int_1^{w\lor 1} F(t,s)c_w(t,s)\ds \leq
      \kappa_2\me^{A(t)}\int_1^{w\lor 1}s^{-\frac{\sharpe}{\sigma\delta_2}}\ds
    \end{equation}
    and similarly
    \begin{equation}\label{eq:H-}
      H(t,c(t,w))^- = \int_{w \wedge 1}^1 F(t,s)c_w(t,s)\ds
      \leq \kappa_2\me^{A(t)}\int_{w\wedge
        1}^1 s^{-\frac{\sharpe}{\sigma\delta_1}}\ds
    \end{equation}

    Suppose first that $\theta < \sigma \delta_2$.  Then $
    H(t,c(t,w))^+ \leq
    \me^{A(t)}\frac{\sigma\delta_2\kappa_2}{\sigma\delta_2-\sharpe}
    w^{1-\frac{\sharpe}{\sigma\delta_2}} \indic{w \ge 1}$.
    Using $W_t=f(t,\lambda(x)Z_t)$ and the estimates in
    \eqref{eq:f_estimates}, we have
    \begin{equation}\label{eq:H+_estimate}
      \begin{split}
        \E H(t,c(t,W_t))^+ & \leq
        \me^{A(t)}\frac{\sigma\delta_2\kappa_2}{\sigma\delta_2-\sharpe}
        \E \left[f(t,\lambda(x)Z_t)^{1-\frac{\sharpe}{\sigma\delta_2}}
          \mathbf{1}_{\lambda(x)Z_t\leq \me^{A(t)}}\right]\\
        &\leq \frac{\sigma\delta_2\kappa_2}{\sigma\delta_2-\sharpe}
        \me^{A(t)\frac{\sigma\delta_2}{\sharpe}}
        \E[(\lambda(x)Z_t)^{1-\frac{\sigma\delta_2}{\sharpe}}]
      \end{split}
    \end{equation}
    and it follows
    that $\E H(t,c(t,W_t))^+<\infty$.

    To estimate the expectation of the integral in time we need a more
    careful analysis.  From Black's equation \eqref{eqn:black}, given
    that $\pits$ is time-homogeneous, we know that
    \begin{equation}\label{eq:beta_explicit}
      \Timefnct(t)=\frac{\sigma^2}{2}\pits_w(w)- r\frac{w}{\pits(w)}
      +\frac{c(t,w)}{\pits(w)},
    \end{equation}
    is a function of $t$ only. Now, depending on whether $\delta_2 =
    \tilde{\delta_2}$ or $\delta_2 = \tilde{\delta_1}$, we let $w\to
    0$ or $w \to \infty$ on the RHS.
    The first term in \eqref{eq:beta_explicit}
    then converges to $\sigma^2 \delta_2/2$, the second term converges
    by l'H\^opital's rule to $r/\delta_2$ and hence also the third
    term converges to some $\kappa_3(t)/\delta_2$, where
    $\kappa_3(t)\geq \kappa_1>0$. We conclude that
    $\beta(t)=\sigma^2\delta_2/2+\kappa_3(t)/\delta_2-r/\delta_2$ and
    \begin{eqnarray} A(t) & = & - \frac{\sharpe}{\sigma}\int_0^t
      \Timefnct(s)ds + \left( \frac{\sharpe^2}{2}-r \right) t
      \nonumber
      \\
      & \leq & \left( -\frac{\sharpe}{\sigma}(\sigma^2\delta_2/2+
        \kappa_1/\delta_2-r/\delta_2)
        +\frac{\sharpe^2}{2}-r\right)t \nonumber \\
      & = & \left( - \frac{\theta \kappa_1}{\sigma \delta_2} - \left(
          1 - \frac{\sharpe}{ \sigma \delta_2} \right) \left(
          \frac{\theta \sigma \delta_2}{2} + r \right) \right)
      t. \label{eq:boundonA}
    \end{eqnarray}
    Using this last estimate in \eqref{eq:H+_estimate} we recover the
    situation in \eqref{eq:crraex_compute}.
    Since by assumption $\sharpe < \sigma \delta_2$, and using the
    representation $Z_t = e^{-rt - \sharpe B_t - \sharpe^2 t/2}$ in
    \eqref{eq:H+_estimate} we have $\E H(t,c(t,W_t))^+ \leq C
    \me^{-\kappa_1 t}$ for a constant $C$, and hence $\E\int_0^\infty
    H(t,c(t,W_t))^+ \dt<\infty$.

    We now turn to the estimates of
    $H(t,c(t,W_t))^-$. In the case
 where $\sharpe < \sigma
    \delta_1$, \eqref{eq:H-} implies that
 $H(t,c(t,w))^- \le
    \kappa_2 \me^{A(t)} \frac{\sigma \delta_1}{\sigma \delta_1 -
      \theta}(1-(w \wedge 1)^{1-\theta/\sigma\delta_1})$. Hence $\E
    H(t,c(t,W_t))^-<\infty$,
 giving also $\E
    |H(t,c(t,W_t))|<\infty$, and for some constant
 $C$,
    $\E\int_0^\infty H(t,c(t,W_t))^- \dt \le C \int_0^\infty
    \me^{A(t)} \, \dt$. Since \eqref{eq:boundonA} implies
 $A(t)/t <
    0$ is bounded away from zero by a constant, then the
 integral is
    finite, and so $\int_0^\infty|H(t,c(t,W_t))|\, \dt <
 \infty$.
    In the case where $\delta_1 < \sharpe/\sigma < \delta_2$ and
    $(\sharpe(1-\delta_2/\delta_1) + \sigma \delta_2) > 0$,
    \eqref{eq:H-} implies $H(t,c(t,w))^- \leq \me^{A(t)}
    \frac{\sigma\delta_1\kappa_2}{\sharpe -
      \sigma\delta_1}w^{1-\frac{\sharpe}{\sigma\delta_1}} \indic{w
      \le
 1}$, and this is decreasing as a function of $w$. Using
    \eqref{eq:f_estimates}, we get
    \begin{equation}\label{eq:W_estimate}
      \begin{split}
        \E H(t,c(t,W_t))^- & \leq
        \me^{A(t)}\frac{\sigma\delta_1\kappa_2}{\sharpe -
          \sigma\delta_1} \E
        \left[f(t,\lambda(x)Z_t)^{1-\frac{\sharpe}{\sigma\delta_1}}
          \mathbf{1}_{\lambda(x)Z_t\geq \me^{A(t)}}\right]\\
        &\leq \frac{\sigma\delta_1\kappa_2}{\sharpe-\sigma\delta_1}
        \me^{A(t)\left(1+\frac{\sigma\delta_2}{\sharpe}-\frac{\delta_2}{\delta_1}\right)}
        \E\left[(\lambda(x)Z_t)^{\frac{\delta_2}{\delta_1}-\frac{\sigma\delta_2}{\sharpe}}\right]
      \end{split}
    \end{equation}
    and it follows that $\E H(t,c(t,W_t))^-<\infty$.
    
    To estimate $\E\int_0^\infty H(t,c(t,W_t))^- \dt$ we use similar
    arguments to those for the positive part, but consider the
    asymptotics of \eqref{eq:beta_explicit} which yield expressions
    involving $\delta_1$. Then we get
    \begin{equation*}
      A(t) \le \left( - \frac{\theta \kappa_1}{\sigma \delta_1} -
        \left( 1 - \frac{\sharpe}{ \sigma \delta_1} \right)
        \left(\frac{ \theta \sigma \delta_1}{2} + r \right) \right)
      t.
    \end{equation*}
    If we write $S:=(\sharpe(1-\delta_2/\delta_1) + \sigma \delta_2)
    >0$, we can compute the expectation as:
    \begin{align*}
      \E H(t,c(t,W_t))^- & \le \frac{\sigma\delta_1\kappa_2
        \lambda(x)^{1-S/\theta}}{\sharpe-\sigma\delta_1}
      \me^{A(t)\frac{S}{\sharpe} - (1-S/\sharpe)(rt+S\sharpe t/2)}\\
      & \le \frac{\sigma\delta_1\kappa_2
        \lambda(x)^{1-S/\theta}}{\sharpe-\sigma\delta_1}
      \me^{-\frac{t}{\sigma \delta_1} \left( \kappa_1 S + (r+\sigma
          \delta_1S/2)(\sigma \delta_1 -S)\right)}.
    \end{align*}
    We now observe that $\sigma \delta_1 - S = (\delta_2 - \delta_1)
    (\sharpe - \sigma \delta_1)/\delta_1 \ge 0$, and we conclude that
    $\E\int_0^\infty H(t,c(t,W_t))^-\dt<\infty$.


    The remaining cases have $\sharpe = \sigma \delta_i$ for some
    $i$. We outline the case where $\sharpe = \sigma \delta_1$ and
    $\delta_1 < \delta_2$, the other cases relying on similar
    calculations. In this case the bounds on $H(t,c(t,W_t))^+$ hold in
    a similar manner to above, while integrating \eqref{eq:H-} and
    using, for $z \ge e^{A(t)}$, $f(t,z) \ge (e^{A(t)} z^{-1})^{\sigma
      \delta_2/\sharpe}$ gives $H(t,c(t,W_t))^- \le \kappa_2
    e^{A(t)}\sigma \delta_2\theta^{-1}\left((A(t))^- + (\ln
      (\lambda(x)Z_t))^+ \right)$. We can derive a lower bound on
    $A(t)$ in a similar manner to \eqref{eq:boundonA}, and note that
    the upper bound also still holds, and so we see that
    $\int_{0}^{\infty} \E H(t,c(t,W_t))^{-} \dt <
    \int_{0}^{\infty}(C_1+C_2 t)e^{-\kappa_1 t} \dt < \infty$ for some
    constants $C_1, C_2$, which gives the required behaviour. In the
    remaining cases where either $\delta_1 < \delta_2 = \sharpe /
    \sigma$, or $\delta_1 = \delta_2 = \sharpe / \sigma$,
    modifications of the above arguments hold.



  \end{proof}

  \begin{remark} \label{rem:suffcondlemma} In Lemma~\ref{lem:globlip}
    we provide two sufficient conditions which relate $\sharpe,
    \sigma$ and the constants $\delta_1, \delta_2$ which are derived
    from $\pi_w(w)$. The simpler necessary condition is to require
    $\sharpe \le \sigma \delta_1$, however in this case, the utility
    function we derive from this consumption/investment pair will
    necessarily be finite in the limit as we let $c \to 0$. To allow
    utility functions which do not display this behaviour, we include
    also the second case. Note however that this second case also
    contains a subset of cases which are easy to check: if $\delta_1 >
    \half\delta_2$, then it is easily confirmed that the second
    necessary condition holds.
  \end{remark}

  \begin{remark} From the proof it is clear that we do not need to
    assume time-homogeneity of $\pits$. Instead we take $\pits(t,w)\in
    \classC^{1,1}$ and assume \eqref{eq:globlip} and
    \eqref{eqn:black}. Then it follows that
  $$\kappa_3(t):=\tilde{\delta_2}\lim_{w\to
    0}\frac{\cts(t,w)}{\pits(t,w)}+\int_1^w
  \frac{\pits_t(t,\xi)}{\pits(t,\xi)^2}\di\xi,\ \kappa_4(t):=
  \tilde{\delta_1}\lim_{w\to
    \infty}\frac{\cts(t,w)}{\pits(t,w)}+\int_1^w
  \frac{\pits_t(t,\xi)}{\pits(t,\xi)^2}\di\xi$$ are well defined. It
  is then enough to assume that $\sharpe/\sigma \le \tilde{\delta}_1
  \land \tilde{\delta}_2$ and further $\int_0^\infty
  \exp(-\int_0^t\kappa_3(u)du)\dt<\infty$ and likewise for
  $\kappa_4(t)$.
\end{remark}

Lemma~\ref{lem:globlip} is particularly useful as it allows us to
construct a wealth of examples of non-linear consumption and
investment pairs with prescribed desired behaviour. We explore now a
method to obtain convex/concave investment and consumptions pairs and
then present a simple parametric family of examples.

Assume that $\pi: \R_+\mapsto \R_+$ is a thrice differentiable,
strictly increasing, concave function such that $\pi(0)=0$ and
\begin{equation}
\label{eq:fconvex}
0 < \pi_w(\infty-)\leq \pi_w(w) < \pi_w(0+) < \infty, \quad w\in (0,\infty).
\end{equation}
Further, let $\chi(w)=(\pi \pi_w)_w(w) =\pi_w(w)^2 + \pi(w)\pi_{ww}(w)$ and assume that there exists $\epsilon>0$ such that
\begin{equation}
\label{eq:glimits}
 - \epsilon^{-1} <  \frac{\sigma^2}{2}\chi(w) <
r - \epsilon, \quad w\in (0,\infty).
\end{equation}
The optimal consumption is given by \eqref{eq:timehom_c} and we assume
that it is time homogenous with $\beta(t)\equiv \beta\geq 0$. We then
have $c(0)=0$ and
\begin{equation*}
\begin{split}
\epsilon \leq \epsilon + \beta \pi_{w}(\infty-) <   c_w(w) &= r + \beta \pi_w(w) - \frac{\sigma^2}{2}\chi(w)
< r + \beta \pi_w(0+) + \epsilon^{-1}.
\end{split}
\end{equation*}
In particular, \eqref{eq:globlip} holds. Provided the Sharpe ratio
satisfies (for example) $\sharpe \le \sigma \pi_w(\infty-)$ we can apply
Lemma~\ref{lem:globlip} and conclude that (i) and (ii) in Theorem
\ref{thm:Dual} hold true.

By hypothesis $\pi$ is concave: the concavity/convexity of $c$ will
depend on the sign of $c_{ww}=\beta \pi_{ww} - \sigma^2
\chi_w/2$. Noting that $\pi_{ww}\leq 0$ we have that if
\begin{equation}\label{eq:cc_condition}
C \pi_{ww}(w) <  \chi_w(w) < 0,\quad w\in (0,\infty),
\end{equation}
for some positive constant $C$ then the choice $\beta = 0$ gives that
$c$ is strictly convex, whereas the
choice $\beta= \sigma^2C/2$ gives that $c$ is strictly concave.

Similarly, we can produce examples with $\pi$ convex. Suppose that
$\pi$ is a thrice differentiable, strictly increasing, convex function
such that $\pi(0)=0$ and $0< \pi_w(0+) \leq \pi_w(w) \leq
\pi_w(\infty-) < \infty$. Then, if we assume again that
\eqref{eq:glimits} holds, we have $c$ such that $c(0)=0$ and
$$\epsilon \leq \epsilon + \beta \pi_{w}(0+) < c_w(w) < r + \beta
\pi_w(\infty-) + \epsilon^{-1}.$$ In conclusion, \eqref{eq:globlip}
holds and if (for example) $\sharpe \le \sigma \pi_w(0+)$ the
assumptions of Lemma \ref{lem:globlip} are satisfied and (i) and (ii)
in Theorem \ref{thm:Dual} follow. If, instead of
\eqref{eq:cc_condition}, we have
\begin{equation}\label{eq:cc_condition2}
  C \pi_{ww}(w) >  \chi_w(w) > 0,\quad w\in (0,\infty),
\end{equation}
then the choice of $\beta=0$ gives a concave consumption while $\beta=\sigma^2C/2$ generates a convex $c$.

\begin{example}\label{ex:non-linconvconc}
  As an example, take in the above $\pi(w) = (\phi w + \psi((1+w)^p -
  1))$ with $\phi, \psi>0$ and $0< p < 1$.  Then $\pi$ is
    concave, $\pi_w(\infty-)=\phi$ and $\pi_w(0+)=\phi + p
  \psi$. Furthermore,
\[ \chi_w(w)=\pi(w) \pi_{www}(w) + 3 \pi_w(w) \pi_{ww}(w) = - \psi(1+w)^{p-3}p(1-p) \Lambda(w) \]
where
\[ \Lambda(w) =
    (1+p) \phi w + (2-p)\psi + 3 {\phi} + 2(1+w)^p \psi (2p-1).
\]
Suppose that the parameters are such that $\Lambda(w)$ is
positive, a simple sufficient condition for which is that $p \geq 1/2$.
Since $\psi>0$ it follows that $\chi$ is decreasing and
\[ \phi^2 = \chi(\infty) \leq \chi(w) \leq \chi(0) = (\phi+ p \psi)^2 \]
so that \eqref{eq:glimits} follows provided $(\phi + p \psi)^2 <
2r/\sigma^2$, a condition we now impose.

We already have $\chi_w<0$ so for \eqref{eq:cc_condition} it suffices to look at the sign of
\begin{equation}
\label{eq:largeC}
C\pi_{ww}- \chi_w = \psi p (1-p) (1+w)^{p-3} [ - C(1+w) + \Lambda(w) ].
\end{equation}
Since $\Lambda$ is bounded above by an affine function of $w$ it
follows easily that this expression can be made negative on
$(0,\infty)$ by choosing $C$ sufficiently large.  Choose the
parameters such that $\Lambda(w)>0$, $(\phi + p \psi)^2 <2r/\sigma^2$
and either $\sharpe \le \sigma \phi$ or both $\sigma \phi < \sharpe \le
\sigma(\phi + p\psi)$ and $\sigma\phi(\phi + p \psi) > \sharpe p
\psi$. Taking $\beta =0$ we obtain an example for which $\pi$ is
concave and $c$ is convex, and conversely, taking $\beta$ sufficiently
large, we obtain an example with both $\pi$ and $c$ concave.

To obtain examples with convex $\pi$ consider now $\phi>0$, $0<p<1$
but $ \psi<0$. Assume also that $\phi + p \psi>0$ and $\phi^2 <
2r/\sigma^2$. Then $\pi$ is increasing, convex with $\phi + p
\psi=\pi_w(0+)<\pi_w(w)<\pi_w(\infty-)=\phi$. Suppose again that the
parameters are such that $\Lambda(w)$ is positive. In fact, given
$\Lambda(0)=3(\phi+p\psi)>0$, $p\leq 1/2$ is a simple sufficient
condition. It follows that $\chi(w)$ is increasing with
\[ (\phi+ p \psi)^2 = \chi(0) \leq \chi(w) \leq \chi(\infty) =
\phi^2 \] so that \eqref{eq:glimits} holds under the condition $\phi^2
< 2r/\sigma^2$. Further, \eqref{eq:largeC} can be made uniformly
positive for large $C$ and hence \eqref{eq:cc_condition2} holds. Lemma
\ref{lem:globlip} applies with either $\sharpe \le \sigma(\phi + p
\psi)$ or both $\sigma(\phi + p\psi) < \sharpe \le \sigma \phi$ and
$\sharpe p \psi > -\sigma\phi(\phi + p \psi)$. We conclude that if we
take $\beta=0$ we obtain an example with convex investment and a
concave consumption. Conversely, if we take $\beta$ sufficiently large
we get an example with both $\pi$ and $c$ convex. A numerical example for this case
is given in Figure \ref{fig:convexconvex}. Note that convexity of $c$
implies DARA by Proposition \ref{prop:riskaversion}.
 \begin{figure}[t]
    \centering
    \includegraphics[width=5.5cm]{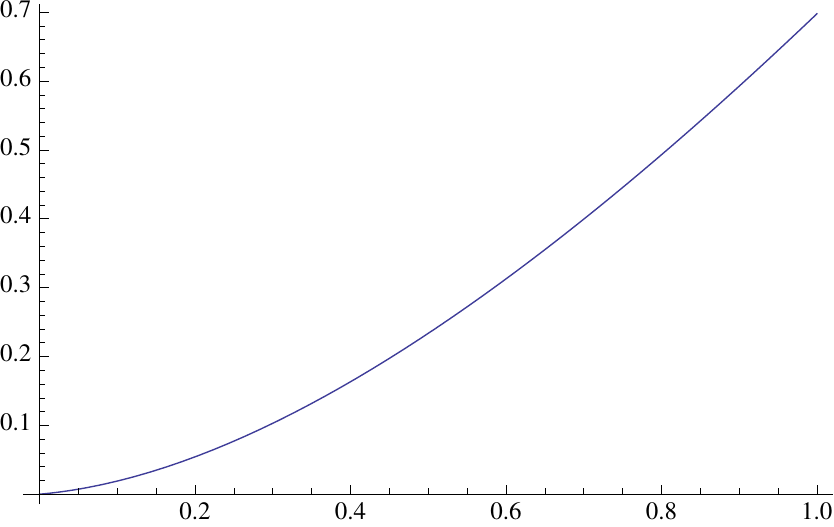}
    \hspace{.5cm}
    \includegraphics[width=5.5cm]{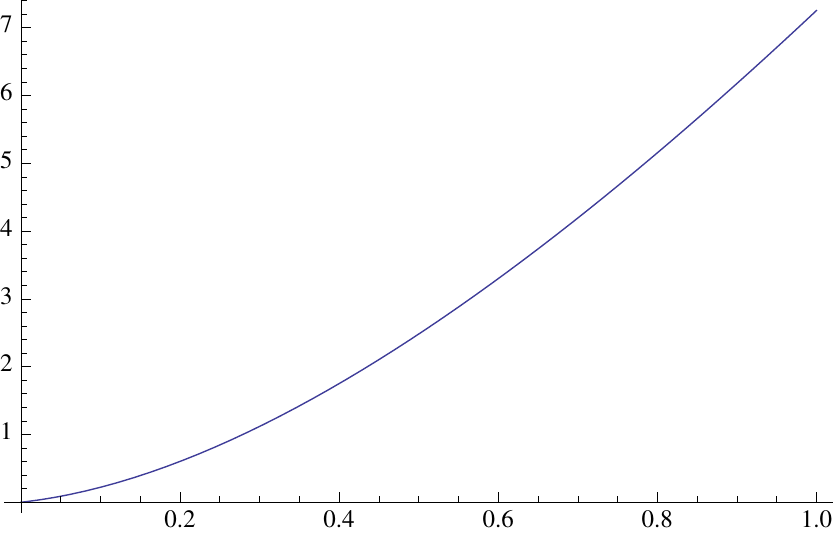}
    \includegraphics[width=5.5cm]{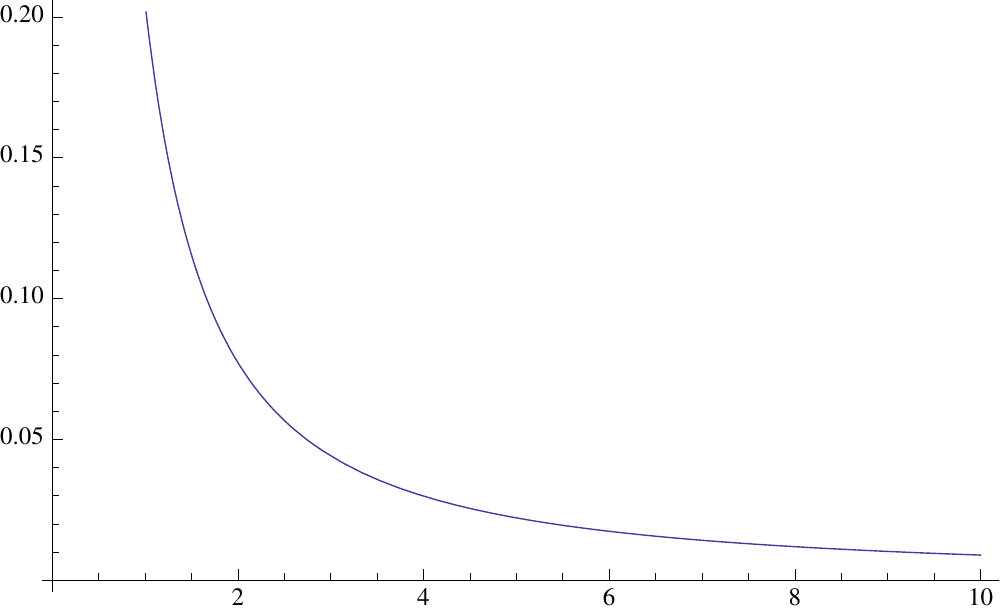}
    \hspace{.5cm}
     \includegraphics[width=5.5cm]{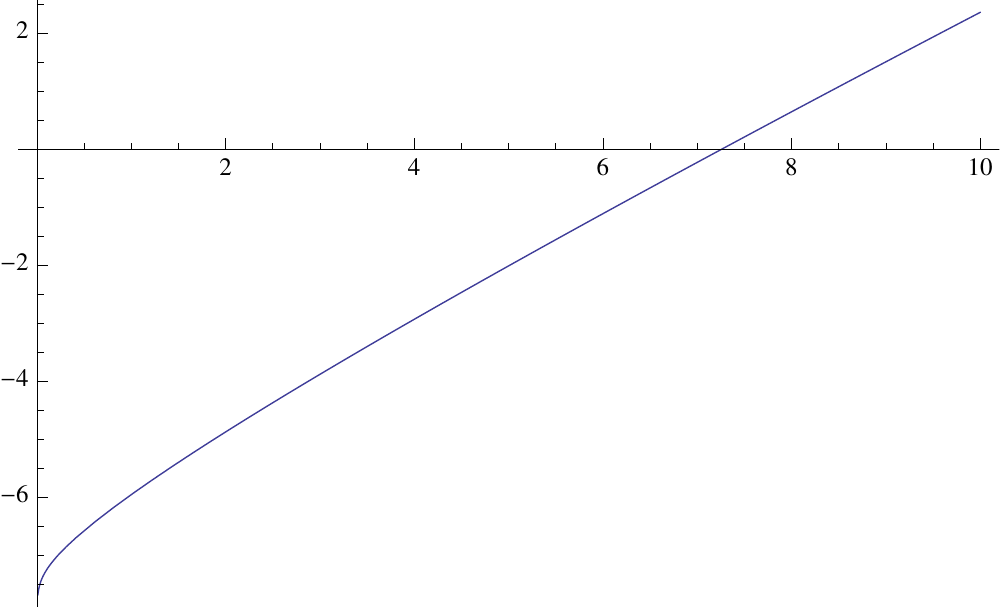}
    \caption{Graphs for Example \ref{ex:non-linconvconc} when $\pi_{ww}>0$.  \emph{Top
        panes}: The investment strategy $\pits(w)$ (left)
      and the corresponding optimal consumption
      $\cts(w)$ (right) for parameters: $r=0.3$,
      $\sharpe=0.026$, $\sigma=0.25$, $\Timefnct=10$, $p=1/30$,
      $\phi=2.1$ and $\psi=-60$.       
      \emph{Bottom panes}: The absolute risk aversion $\rho(t,c)=\rho(c)$
      (left) inferred from these actions and a compatible utility
      function $u(0.1,c)$ (right).
\label{fig:convexconvex}
}
\end{figure}

Our general approach easily allows us to obtain admissible sets of
parameters with additional convexity and concavity properties. Note,
however, that even when the arguments for the convexity/concavity fail,
it is still straightforward to write down expressions for $c_w$ and analyse it explicitly. In particular, when $\psi>0$ we see that 
 $\Timefnct > \frac{\sigma^2}{2\phi}(\phi+p\psi)^2$ guarantees
  that $c_w$ is bounded and bounded away from zero. As above, Lemma
\ref{lem:globlip} applies with either $\sharpe \le \sigma \phi$ or both $\sigma \phi < \sharpe \le
\sigma(\phi + p\psi)$ and $\sigma\phi(\phi + p \psi) > \sharpe p
\psi$. A numerical example which satisfies these conditions but for which
$\Lambda(w)$ goes negative is given in Figure \ref{fig:concaveconcave}.
It also features a risk aversion which is first decreasing,
then increasing and then decreasing again.
  \begin{figure}[ht]
    \centering
    \includegraphics[width=5.5cm]{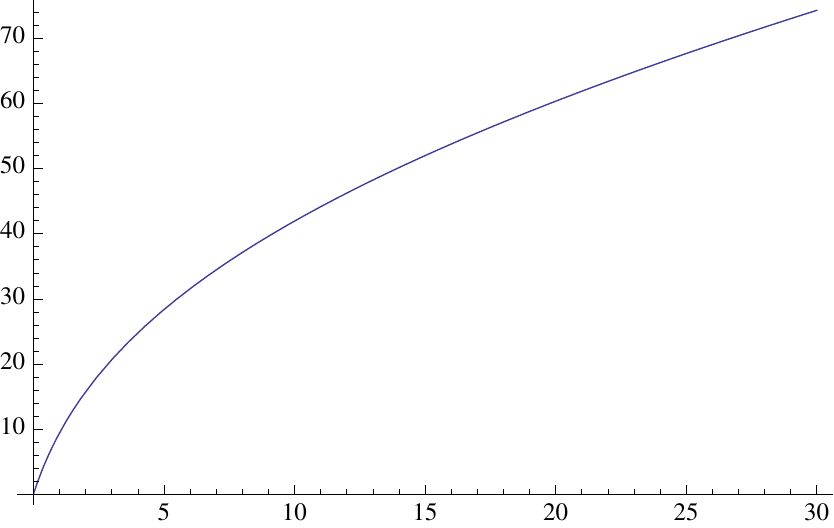}
    \hspace{.5cm}
    \includegraphics[width=5.5cm]{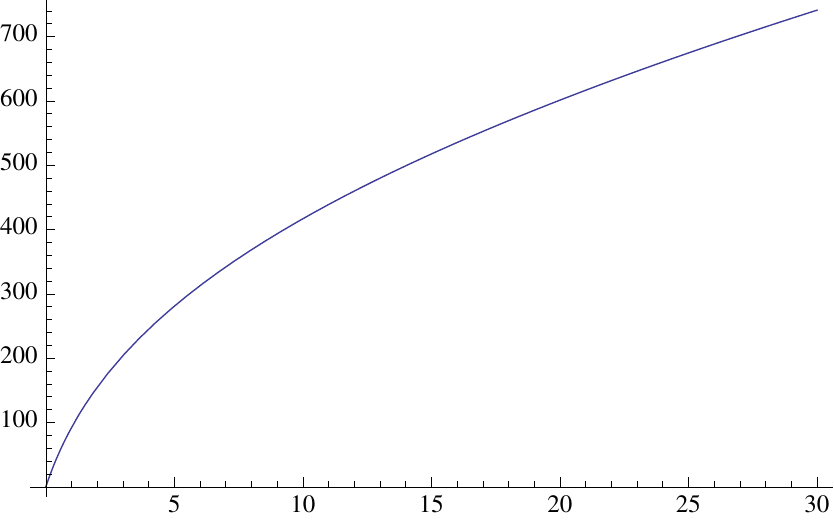}
    \includegraphics[width=5.5cm]{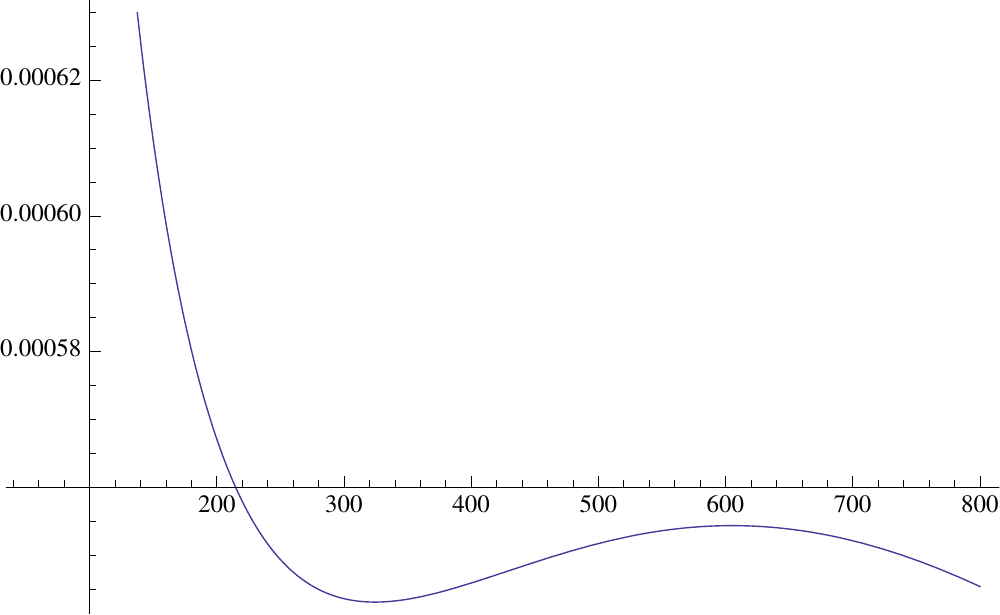}
    \hspace{.5cm}
    \includegraphics[width=5.5cm]{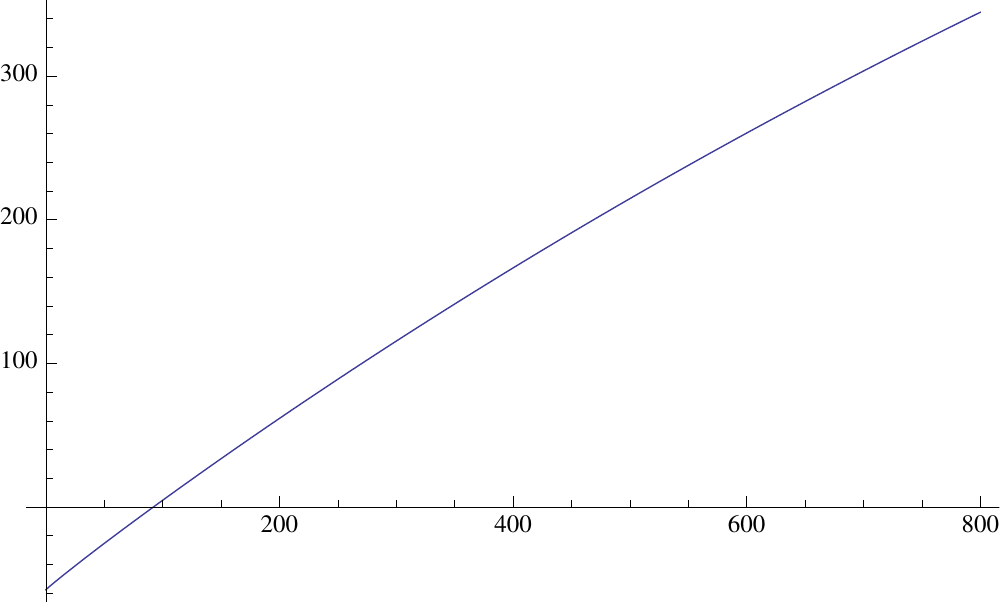}
    \caption{Graphs for Example \ref{ex:non-linconvconc} when $\pi_{ww}<0$.  \emph{Top
        panes}: The investment strategy $\pits(w)$ (left)
      and the corresponding optimal consumption
      $\cts(w)$ (right) for parameters: $r=0.05$,
      $\sharpe=0.13$, $\sigma=0.25$, $\Timefnct=10$, $p=1/5$,
      $\phi=0.5$ and $\psi=60$. 
      \emph{Bottom panes}: The absolute risk aversion $\rho(t,c)=\rho(c)$
      (left) inferred from these actions and a compatible utility
      function $u(0.1,c)$ (right).
\label{fig:concaveconcave}
}
\end{figure}

\end{example}

\begin{example}\label{ex:convex_c}
We present another example where consumption has a simple convex expresion in wealth. Consider
  \begin{equation}\label{eq:pi_ex3}
    \pi(w)=\frac{2}{\sigma}\sqrt{\frac{r-\kappa}{2}w^2+\alpha w +
      \frac{\alpha}{a}(\me^{-aw}-1)},
  \end{equation}
  where we take $r>\kappa>0$, $\alpha,a>0$ with $\kappa>a\alpha$. Clearly $\pi(w)$ is an
  increasing function of $w$ and
  \begin{equation*}
    \delta_2 = \pi_w(0+)=\frac{\sqrt{2}}{\sigma}\sqrt{r -\kappa+\alpha a},\quad
    \delta_1 =  \pi_w(\infty-) =\frac{\sqrt{2}}{\sigma}\sqrt{r -\kappa}.
  \end{equation*}
  From \eqref{eq:timehom_c}, with $\Timefnct(t)\equiv 0$, we recover
  the optimal consumption as
  \begin{equation}
    \label{eq:c_ex3}
    \cts(w)= \kappa w + \alpha\left(\me^{-aw}-1\right)
  \end{equation}
  which is an increasing convex function with $c_{ww}(w)=a^2\alpha
  \me^{-aw}$.  Sufficient conditions for $\pits,\cts$ to satisfy the
  assumptions of Lemma~\ref{lem:globlip} are then concavity of $\pits$
  (from which we deduce $\delta_1 \le \pits_w(w) \le \delta_2$) and
  either $\sharpe \le \sqrt{2(r-\kappa)}$ or $\sqrt{2(r-\kappa)}
  <\sharpe \le \sqrt{2(r-\kappa+a\alpha)}$ and $\sharpe +
  \sqrt{2(r-\kappa+a\alpha)}) >\sharpe\sqrt{1+a
    \alpha/(r-\kappa)}$. We will
  verify the concavity condition numerically for the cases of interest
  presented in Figure~\ref{fig:ex_convex}.  Note that in this example
  $c$ is convex and $\pi$ is increasing so
  Proposition~\ref{prop:riskaversion} implies that the agent employing
  these actions necessarily has a decreasing absolute risk aversion.
  \begin{figure}
    \centering
    \includegraphics[width=5.5cm]{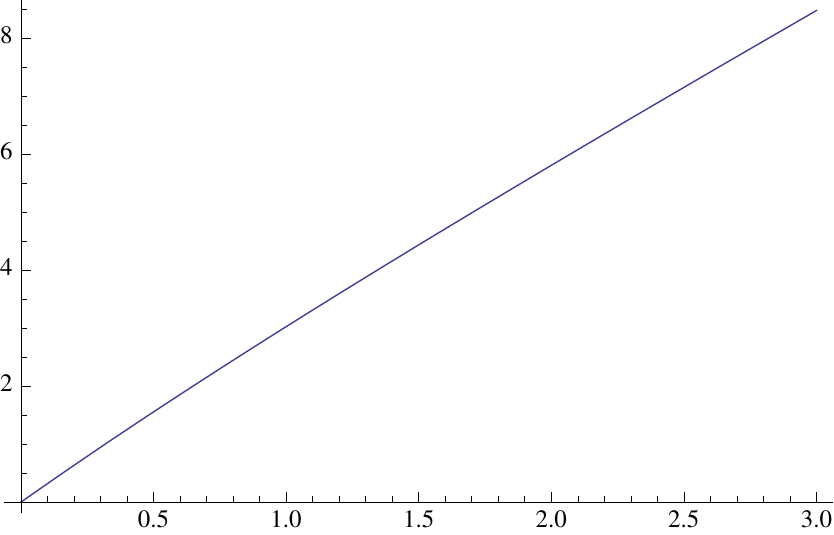}
    \hspace{.5cm}
    \includegraphics[width=5.5cm]{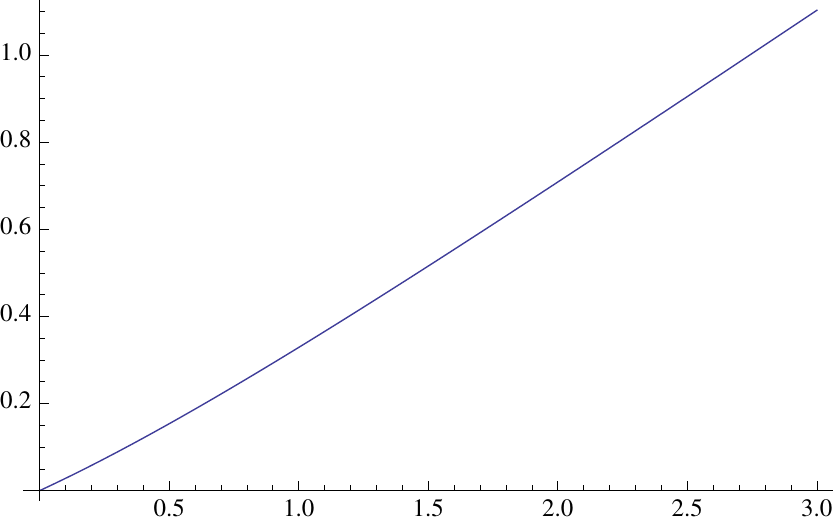}
    \includegraphics[width=5.5cm]{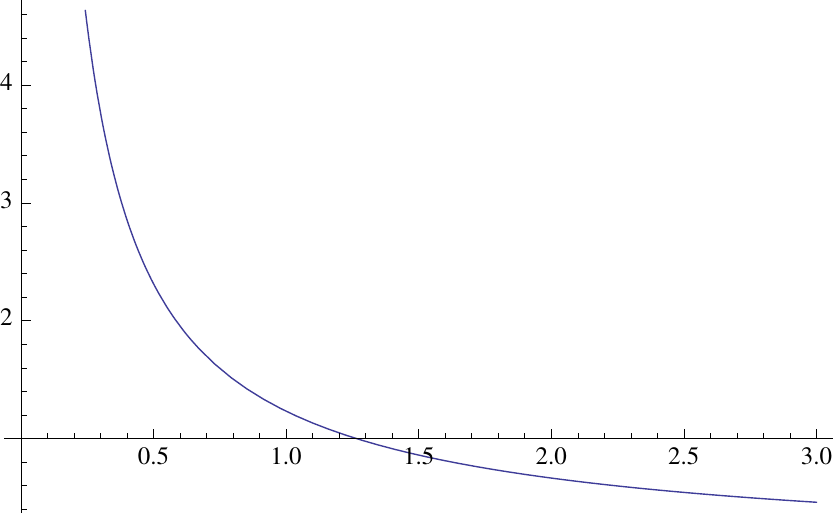}
    \hspace{.5cm}
    \includegraphics[width=5.5cm]{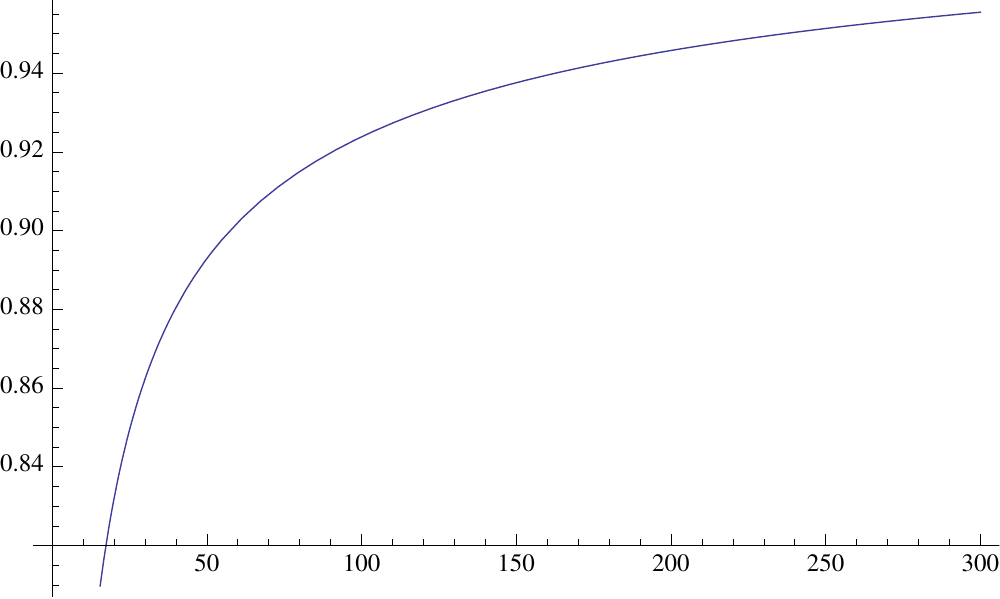}
     \caption{Graphs for Example \ref{ex:convex_c}.
      \emph{Top panes}: The investment strategy $\pits(w)$ (left) in
      \eqref{eq:pi_ex3} and the
      corresponding optimal consumption $\cts(w)$ (right) in
      \eqref{eq:c_ex3} for parameters:
      $\kappa = 0.4, \sigma = 0.25, r = 0.6, \alpha = 0.1, a= 1.25, \sharpe = 0.95$.
      \emph{Bottom panes}: The absolute risk aversion $\rho(t,c)=\rho(c)$
      (left) inferred from these actions and a compatible utility
      function $u(0.1,c)$ (right).
      \label{fig:ex_convex}
    }
  \end{figure}
\end{example}

\subsection{Extensions to Theorem~\ref{thm:Dual}} \label{sec:bdd}

Our goal in Theorem~\ref{thm:Dual} was to present an if and only if statement
of our main result in the regular case. In this section we talk about a few of
the extensions which are possible.

Firstly, note that in $(i)$ of Theorem~\ref{thm:Dual} we only assume
regularity on $u$ and in $(ii)$ we only assume regularity on
$\cts,\pits$. With a standing assumption that $\pits$ is continuously
differentiable in both arguments the equivalence holds under {weaker
  definitions of regularity}.  In $(i)$ it is then enough to require
that $u(t,\cdot)$ is once (and not twice) continuously differentiable
and we can drop H\"older continuity of $I_z$. In $(ii)$ we then need
that $\cts$ is jointly continuous instead of differentiability in $w$
and H\"older continuity of $c_w$. The key point is that we need to
guarantee the existence of $g_{tz}$ in the second part of the proof of
the theorem, and this can be done either with the assumptions of the
theorem, or with weaker conditions if $\pits$ is assumed to be
continuously differentiable.

Another possible extension is to drop the Inada conditions on $u$. The
implications of not requiring $u_c(t,0)=\infty$ --- in particular,
consumption may be zero for an interval of positive wealths to the
right of zero --- is considered in \citet{HeHuang:94}.

Instead of pursuing this idea, for the rest of this section we will
consider what happens if agent's utility includes a satiation point,
or in the inverse case where we start with $c$ and $\pi$, if the
consumption is a bounded function of wealth.
This may or may not imply that agent's wealth is bounded (and below we 
give examples which cover the two cases), but in the
case where the wealth is bounded, there are similarities with the
`maximal' wealth path $\ol w(t)$, which arose in the deterministic
setting.

Implicit in the definitions and results of
Section~\ref{sec:main_stoch} (and in \citet{Black:68},
\citet{CoxLeland:00}, \citet{HeHuang:94}) is the idea that the agent
follows a strategy such that his wealth is unbounded in $\omega$ for
each $t$. The result below considers the case where the consumption
$c(t,w)$ may be bounded above, and then $W^x_t\leq \ol w(t)$ a.s.\
where $\ol w(t) = \me^{rt} \int_t^\infty\me^{-rs} c(s,\infty) \ds$
which may be finite or infinite.  In the former case, the agent
consumes and invests in such a way that his wealth is kept below
$\ol{w}(t)$ with probability one. We relate this explicitly to the
properties of the utility function.

The following definition relaxes the notions from Definitions
\ref{def:utility} and \ref{def:regularpic}.
\begin{definition}\label{def:relaxed_ut_pic}
  (i) We say that a function $u:[0,\infty)^2\to [-\infty,\infty)$ is a
  \emph{regular utility function} if for any $t\geq 0$ there exists
  $\ol{c}(t)>0$ such that $u(t,\cdot)$ is twice continuously
  differentiable, strictly concave and increasing on $\{(t,c)\in
  (0,\infty)^2: c< \ol{c}(t)\}$, satisfies the Inada condition:
  $u_c(t,0+)=\infty$ and $u_c(t,\ol{c}(t)-)=0$, and $u(t,c)$ is
  constant for $c \geq \ol{c}(t)$. Further, $I_z$ is locally H\"older
  continuous on $(0,\infty)^2$.

(ii) We say that $(c,\pi)$ is a \emph{regular consumption/investment
pair} if 
\begin{itemize}
\item for each $t\geq 0$, $\cts(t,0)=0$, $\cts(t,\cdot)$ is increasing
  and we have $\ol w(t):= \me^{rt}\int_t^\infty \me^{-rs} c(s,\infty)\ds>0$.
  Further, $c(t,\cdot)$ is strictly increasing and differentiable on
  $[0,\ol w(t))$ and constant and equal to $c(t,\ol w(t)-)$ on $[\ol
  w(t),\infty)$.  Finally, $c_w(t,w)$ is locally H\"older continuous
  on $\{(t,w)\in (0,\infty)^2: w<\ol w(t)\}$.
\item for each $t\geq 0$, $\pits(t,0)=0$, $\pits(t,\cdot)$ is strictly
  positive and continuously differentiable in both arguments on
  $\{(t,w)\in (0,\infty)^2: w<\ol w(t)\}$. Further, $\int_{0+} \di \xi
  / \pi(t,\xi)= \infty = \int^{\ol{w}(t)} \di \xi / \pi(t,\xi)$ and
  $\pi(t,w)=0$ for $w \geq \ol w(t)$ when $\ol w(t)<\infty$.
\end{itemize}
Finally, $c$, $\pi$ are such that the SDE \eqref{eq:wealth_dynamics} has a
strong solution for $x < \ol{w}(0)$.
\end{definition}

In addition, the definitions of $\beta$, $A$, $H$ and $F$ may need to be 
modified. If $\ol w(t)>\xi>0$ for all $t\geq 0$ then
it suffices to replace $1$ in the lower bound of the integration in
\eqref{eqn:black} and \eqref{eq:def_F} by $\xi$. More generally,
when $(\pi,c)$ is a regular
consumption/investment pair as given by the
Definition~\ref{def:relaxed_ut_pic}, we have that $\ol w(t)$ is
continuous and strictly positive and then there exists a smooth function
$w_0(t)$ such that $0 < w_0(t) < \ol w(t)$. Thus we can replace the 
bottom
limit in the integrals in \eqref{eqn:black} and \eqref{eq:def_F} with
$w_0(t)$, and replace $A(t)$ in \eqref{eq:def_F} with
\begin{equation*}
  A_0(t) = - \frac{\sharpe}{\sigma} \int_0^t
  \Timefnct(s) \ds + \left(\frac{\sharpe^2}{2} - r\right)t - \int_0^t
  \frac{\theta w_0'(s)}{\sigma  \pi(s,w_0(s))} \ds.
\end{equation*}

Similarly, if $\ol{c}(t)$ is not bounded below by 1, then we replace the 
lower limit in the definition of $H$ with $c_0(t)$ where $c_0(t)< 
\ol{c}(t)$ is some strictly positive function. Then
$H(t,c)= \int_{c_0(t)}^{c} F(t,Y(t,b))db$.

\begin{theorem}
  \label{thm:DualS}
  For any $x>0$, the following are equivalent:
  \begin{enumerate}
  \item $c(t,W^{x}_t)$ and $\pi(t,W^{x}_t)$ achieve a finite maximum
    in the problem \eqref{eq:stocutility}
    for a regular utility function $u$ of Definition \ref{def:relaxed_ut_pic} for which \eqref{eq:ExistsLM} holds, and where $c,\pi:[0,\infty)^2 \to [0,\infty)$ are such that if $\ol w(t):= \inf\{w>0: \pi(t,w)=0\}<\infty$ then
    $\pi(t,w)=0$, $c(t,w)=c(t,\ol w(t))$ for $w\geq \ol w(t)$.
  \item $c(t,w),\pi(t,w)$ are a regular consumption/investment pair of Definition \ref{def:relaxed_ut_pic} which satisfy \eqref{eqn:black} on $\{(t,w)\in (0,\infty)^2: w<\ol w(t)\}$,
\eqref{eq:Budget} holds and for some $0<x_0\leq x$,
$\E[|H(t,c(t,W^{x_0}_t))|]<\infty$ for
    almost all $t \ge 0$ and $\int_0^\infty \E[ H(t,c(t,W^{x_0}_t)) -
    h(t) ]^+ \dt<\infty$, where $h(t)=\E[H(t,c(t,W^{x_0}_t))]$.
  \end{enumerate}
  Moreover, we then have $u_c(t,c)=H_c(t,c)$, and in
  (i) one may take $u(t,c)=H(t,c)-h(t)$ for $c \leq \ol{c}(t)$. 
\end{theorem}

\begin{remark} The theorem holds if in the definition of a regular 
consumption/investment pair $\ol w(t)$ is just some function and we do 
not impose the consistency condition that $\ol w(t):= 
\me^{rt}\int_t^\infty \me^{-rs} c(s,\infty)\ds$. This condition is in 
fact implied by Black's equation \eqref{eqn:black} which gives 
$W_t^x=f(t,F(0,x)Z_t)$ and by the budget equation \eqref{eq:Budget}, as 
is clear from the proof of $(i)\Rightarrow (ii)$. 
\end{remark}

\begin{proof} The proof is almost identical to the proof of
  Theorem~\ref{thm:Dual}. In the first part, when showing
  $(ii)\Rightarrow (i)$, observe that $u(t,c)=H(t,c)-h(t)$ is strictly
  increasing and concave on $[0,\ol c(t))$ and constant on $[\ol
  c(t),\infty)$, where $\ol c(t)= c(t,\infty)=c(t,\ol w(t))$. We have
  $u_c(t,0)=F(t,Y(t,0))=F(t,0)=\infty$, $u_c(t,\ol c(t)) =F(t,Y(t,\ol
  c(t))) =F(t,\ol w(t))=0$ and $u_{cc}(t,l)$ is continuous on $(0, \ol
  c(t))$. In addition, $I(t,z) = c(t,f(t,z))$ and $f$ is well defined
  and $\classC^{1,2}$ on $(0,\infty)^2$.  In consequence, $I_z(t,z)$
  is locally H\"older continuous and $u$ is a regular utility function
  of Definition \ref{def:relaxed_ut_pic}. The modified definition of
  $F(t,w)$ (note in particular that \eqref{wFa}--\eqref{wFc} still
  hold) is important in \eqref{eq:Ftused}, where the additional terms
  in $A_0(t)$ cancel with the extra term in $F_t$ arising from the new
  definition.
    
    For the second part of the proof, the implication $(i)\Rightarrow (ii)$, recall that now $\ol c(t)$ is defined from $u$ in Definition \ref{def:relaxed_ut_pic}. When we show that $c(t,W^x_t)=I(t,\lambda Z_t)$ this implies $c(t,W_t^x)<\ol c(t)$. 
    Then, from the representation of $g$ as the conditional expectation, letting $Z_t\to 0$ and using the Dominated Convergence Theorem, we obtain that $\ol{c}$ and $\ol{w}$ are related by $\ol w(t) = \me^{rt} \int_t^\infty\me^{-rs} \ol{c}(s) \ds$. 
We conclude that $c(t,0)=0$, $c(t,\cdot)$ is strictly increasing on $(0, g(t,0))$, and $c(t,g(t,0))=\ol c(t)$. Further, $c_w(t,w)$ is locally H\"older
    continuous on $\{(t,w)\in (0,\infty)^2: w< g(t,0)\}$. As previously, \eqref{pidef}--\eqref{Cdef} hold with $g$ instead of $f$, for all $t,z>0$. It follows that $\pi(t,0)=0$, and $\pi(t,w)>0$ for $0<w<g(t,0)$. Moreover, if $g(t,0)<\infty$ then $g_z(t,0)>-\infty$ and $\pi(t,g(t,0))=0$. It follows from the assumed properties of $c,\pi$ that $g(t,0)=\ol w(t)$ and $\ol c(t)=c(t,\infty)$.
    Similarly, we conclude that \eqref{eqn:black} holds for $\{(t,w)\in (0,\infty)^2: w<\ol w(t)\}$. Finally, the last change is that now 
    $$\tau=\inf\{s: Z_s\notin (F(s,\ol w(s))/F(0,x),F(s,0)/F(0,x))\}=\inf\{s: \tilde{W}_s \in
    \{0,\ol w(s)\}\}$$
and $0<W_t^{x}<\ol w(t)$ a.s.\ implies that $\tau=\infty${;} i.e.\ $F(t,0)=\infty$ and
    $F(t,\ol w(t))=0$.
\end{proof}
\begin{example}\label{ex:bounded_wealth}
  We now present an example where the situation as above holds: both the agent's consumption and her wealth process are bounded. Consider the time-homogeneous setting with
  $\pits(w) = \max\{w(1-w),0\}$ and
  \begin{equation*}
    c(w) = w \left(r-\half \sigma^2 + \beta\right) + w^2
    \left(\frac{3}{2} \sigma^2 - \beta\right) - w^3 \sigma^2,\quad w\in (0,1),
  \end{equation*}
  with $c(w)=r$ for $w\geq 1$. We shall suppose that $r>\half
  \sigma^2$ and $|\beta| < r-\half \sigma^2$. Observe that $c(0) = 0$,
  $c$ is strictly increasing on $(0,1)$ with derivative bounded away
  from zero and infinity and $c$ is continuous on
  $(0,\infty)$. Further, we have $\ol w(t) \equiv 1=\int_t^\infty
  \me^{r(t-s)}c(1)\ds$.  In particular, $(c,\pi)$ are a regular
  consumption/investment pair which satisfy \eqref{eqn:black} as
  described above. Moreover, since $W_t$ is bounded above, similar
  arguments to those used at the start of the proof of
  Lemma~\ref{lem:globlip} can be used to deduce
  \eqref{eq:Budget}. Finally, we can also check the integrability
  conditions of {\it (ii)} of Theorem~\ref{thm:DualS}. Then we can
  conclude that there exists a regular utility function for which this
  consumption and investment are optimal.

  First note that we have
  \begin{equation*}
    F(t,w) = \me^{A(t)} \left( \frac{w}{1-w}\right)^{-\theta/\sigma},
  \end{equation*}
  so that
  \begin{equation} \label{eq:Fbounds}
    \half w^{-\theta/\sigma} \indic{w < \delta} \le \me^{-A(t)} F(t,w)
    \le w^{-\theta/\sigma}
  \end{equation}
  for $w \in (0,1)$, and for some $\delta < \half$. Again, as in the
  proof of Lemma~\ref{lem:globlip}, we can write
  \begin{equation*}
    H(t,c(t,w)) = \int_{\half}^w F(t,\tilde{w}) c_w(t,\tilde{w})\di \tilde{w}.
  \end{equation*}
  Since $c_w$ is bounded from above and below for $w \in (0,1)$, then
  we deduce from \eqref{eq:Fbounds} that $H(t,c(t,w))^+ \le \kappa_1
  \me^{A(t)}$ for some constant $\kappa_1 >0$, and $H(t,c(t,w))^- \le
  \kappa_2 \me^{A(t)}$ when $\theta < \sigma$, and $H(t,c(t,w))^- \le
  \kappa_3 \me^{A(t)} w^{1-\theta/\sigma}$ when $\theta >\sigma$, for
  some $\kappa_2, \kappa_3 >0$. (We exclude the case $\theta =
  \sigma$). Writing $f(t,z)$ for the inverse of $F(t,w)$ and using
  \eqref{eq:Fbounds}, we get $f(t,Z_t) \le \me^{A(t) \sigma/\theta}
  (Z_t \vee K)^{-\sigma/\theta}$, for some $K>0$, and we conclude that
  the desired integrability conditions hold provided
  \begin{equation*}
    \int_0^\infty \me^{A(t)\sigma/\theta}\left( \E
      \left[Z_t^{1-\sigma/\theta}\right] + K^{1-\sigma/\theta}\right) \di t < \infty.
  \end{equation*}
  Calculations similar to those in the proof of
  Lemma~\ref{lem:globlip} show that this will hold whenever:
  \begin{equation*}
    \beta - \sigma \theta + r - \half \sigma^2 > 0 \text{ and } A(1) =
    -\frac{\theta}{\sigma} \beta + \half \theta^2-r < 0.
  \end{equation*}
  It is now clear that there are non-trivial parameter choices for
  which this example satisfies the conditions of
  Theorem~\ref{thm:DualS}, and therefore, such that there exists a
  utility function $u$ for which these are the optimal
  investment/consumption pair.
\begin{figure}
    \centering
    \includegraphics[width=5.5cm]{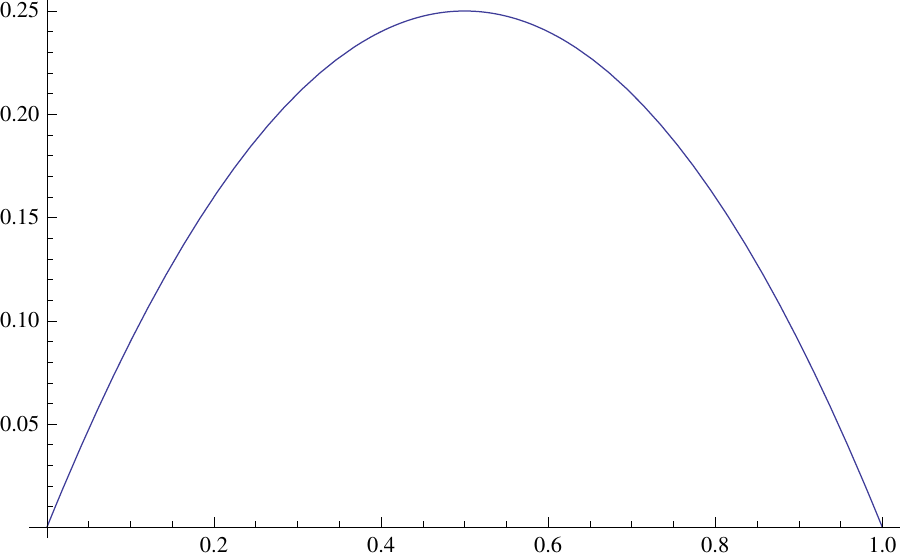}
    \hspace{.5cm}
    \includegraphics[width=5.5cm]{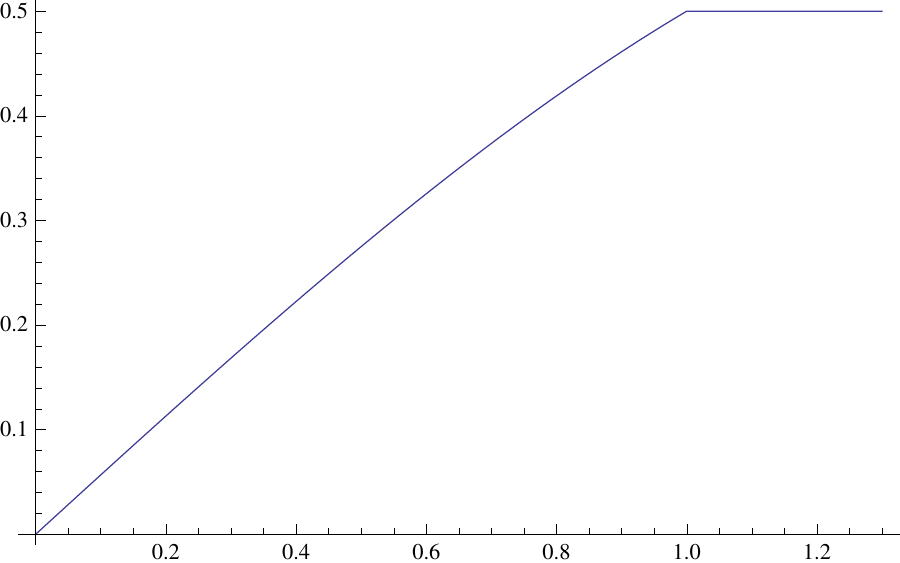}
    \includegraphics[width=5.5cm]{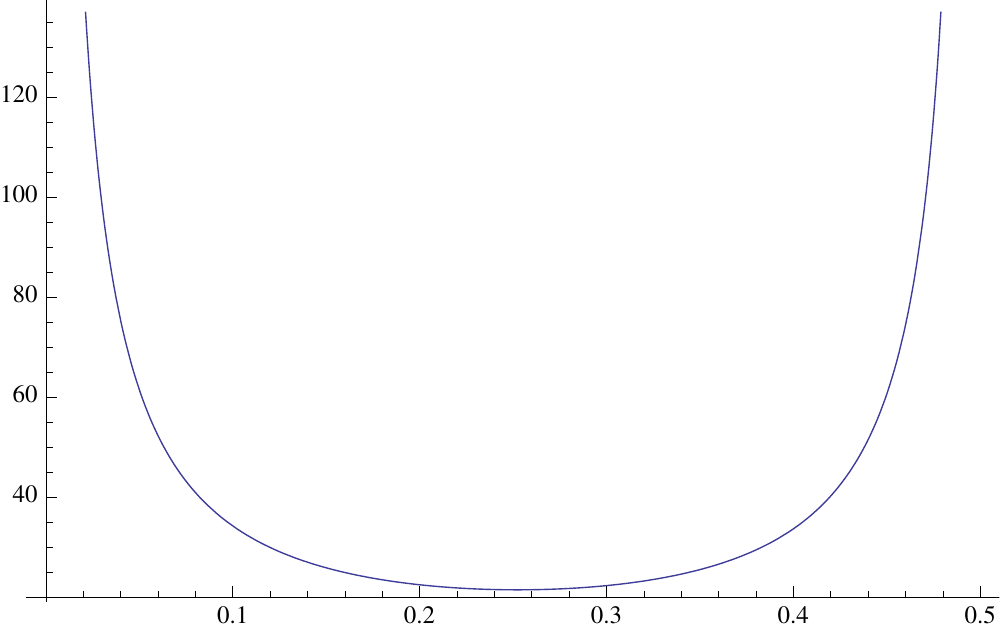}
    \hspace{.5cm}
    \includegraphics[width=5.5cm]{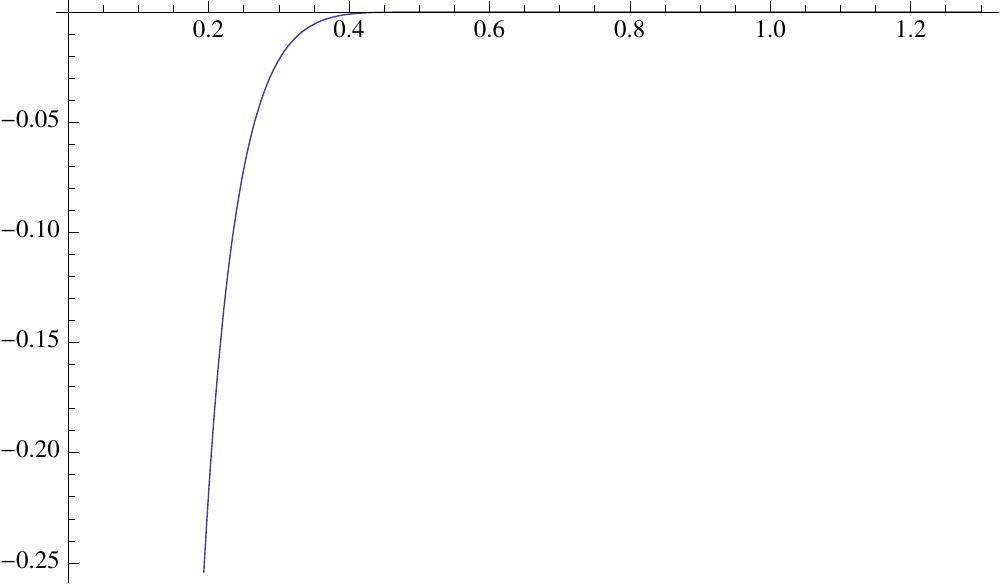}
  \caption{Graphs for Example \ref{ex:bounded_wealth}.
    \emph{Top panes}: The investment strategy $\pits(w)=w(1-w)\lor 0$ (left) and the
    corresponding optimal consumption $\cts(w)$ (right) for parameters:
    $r=0.5$, $\sharpe=0.7$, $\sigma=0.25$, $\beta=0.1$.
    \emph{Bottom panes}: The absolute risk aversion $\rho(1,c)$ (left) inferred from these actions and a compatible utility function $u(1,c)$ (right) which is constant on $[r,\infty)$.
    \label{fig:ex_bounded}
  }
\end{figure}
\end{example}

\begin{example}\label{ex:bounded_cons}
Finally, we present an example in which the consumption is bounded 
while the agent's wealth is unbounded. 
Consider the time-homogeneous setting with $\pits(w)=1-\me^{-w}$ and
$$c(w)=\left(\beta-\frac{\sigma^2}{2}\me^{-w}\right)\pits(w),\quad w\geq 0.$$
We assume $r=0$, $\theta<\sigma$ and $\beta>
\sigma^2$. It follows that $c$ is an increasing
function with $c(0)=0$ and $c(\infty)=\beta$.  Further, $\pi,c$
satisfy Black's equation \eqref{eqn:black} as $c$ is given by
\eqref{eq:timehom_c}. Explicit computations yield
$$F(t,w)=\me^{A(t)}\left(\me^{w}-1\right)^{-\sharpe/\sigma},\quad f(t,z)=\log\left(\me^{A(t)\sigma/\sharpe}z^{-\sigma/\sharpe}+1\right),$$
where we used $\log 2$ instead of $1$ as the lower bound of
integration in \eqref{eq:def_F}. Then $c,\pi$
are a regular consumption/investment pair of Definition
\ref{def:relaxed_ut_pic} with $\ol w(t)=\infty=\int_t^\infty
c(\infty)\ds$, and we note that $F(t,\infty)=0$ as required.
Similarly to Example \ref{ex:bounded_wealth} above, it is easy to see
that $H(t,c(t,w))\leq \kappa \me^{A(t)}$ when
$\sharpe<\sigma$. Likewise, with arguments akin to that in the proof
of Lemma~\ref{lem:globlip}, we obtain that $W_t^x\to 0$ and
\eqref{eq:Budget} is equivalent to showing that $\E^\Q[W_t^x]\to 0$. To
verify this we use Remark \ref{rk:wealth_expression} and compute
\begin{equation*}
\begin{split}
\E^\Q[W_t^x]& = \E^\Q[f(t, F(0,x)Z_t)]=\E^\Q\left[\log\left(\me^{\sigma (B_t+\sharpe t) -\beta t}(\me^{x}-1)+1\right)\right]\\
& \leq \log \left(\E^\Q\left[\me^{\sigma (B_t+\sharpe t) -\beta t}(\me^{x}-1)+1\right]\right)
=\log \left(\me^{-(\beta-\sigma^2/2)t}(\me^{x}-1)+1\right)\\ & \to 0,
\end{split}
\end{equation*}
as $t\to \infty$, and where we used Jensen's inequality, the fact that $B_t+\sharpe t$ is a $\Q$--Brownian motion and the assumption $\beta>\sigma^2/2$.
Finally, note that $\beta\geq \sigma^2/2$ and $\sharpe<\sigma$ together imply $A(1)=
-\frac{\theta}{\sigma} \beta + \half \theta^2<0$ and we conclude that all the assumptions in $(ii)$  in Theorem \ref{thm:DualS} are satisfied.
\begin{figure}
    \centering
   \includegraphics[width=5.5cm]{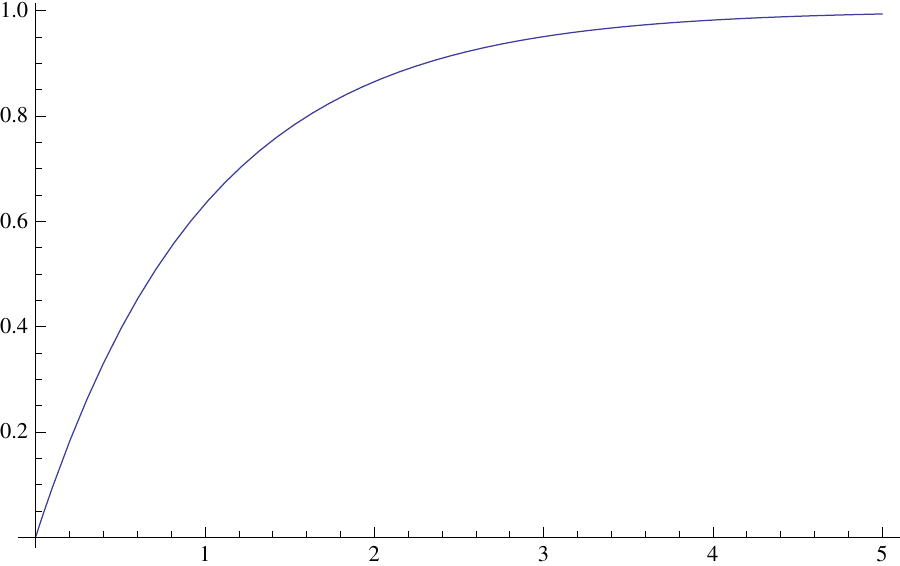}
    \hspace{.5cm}
    \includegraphics[width=5.5cm]{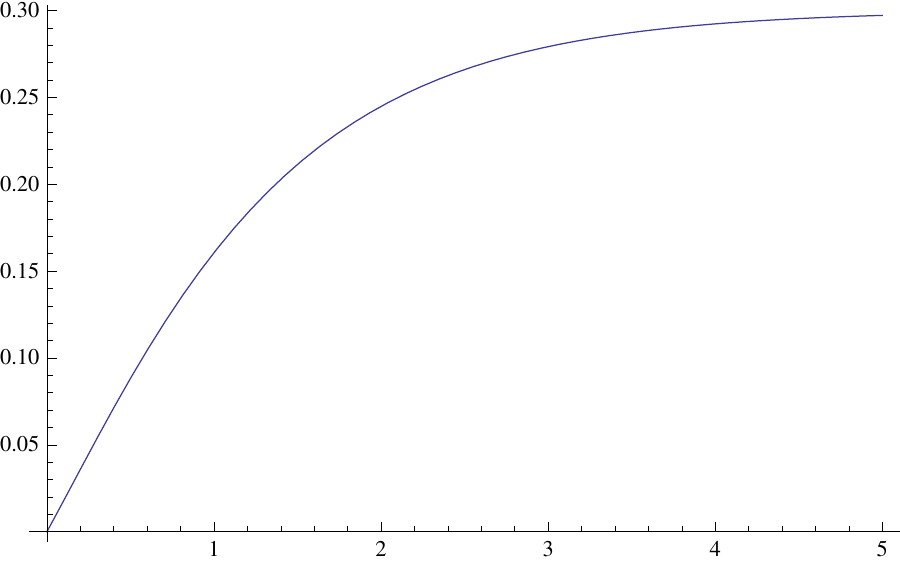}
   \includegraphics[width=5.5cm]{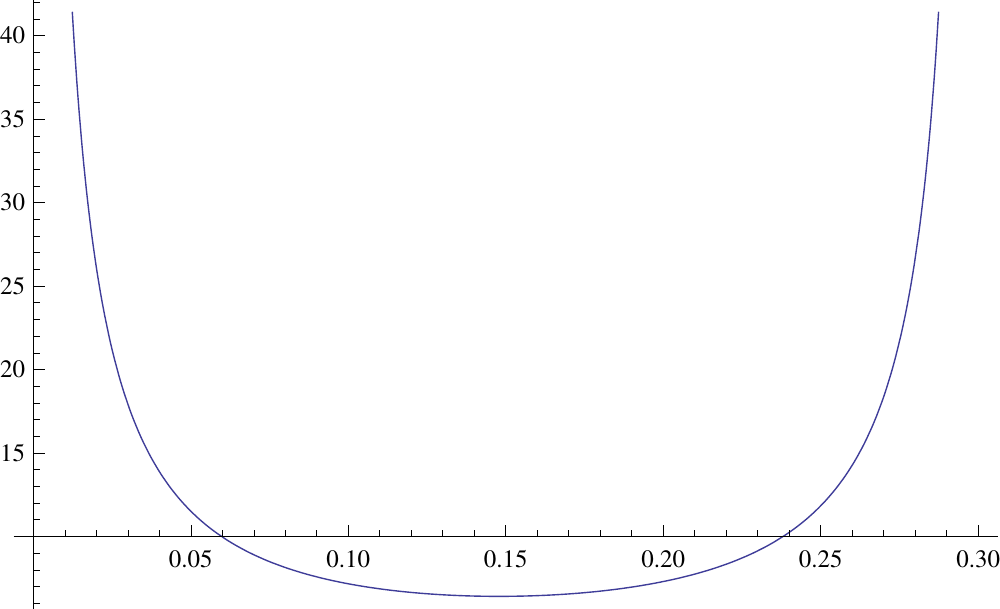}
    \hspace{.5cm}
    \includegraphics[width=5.5cm]{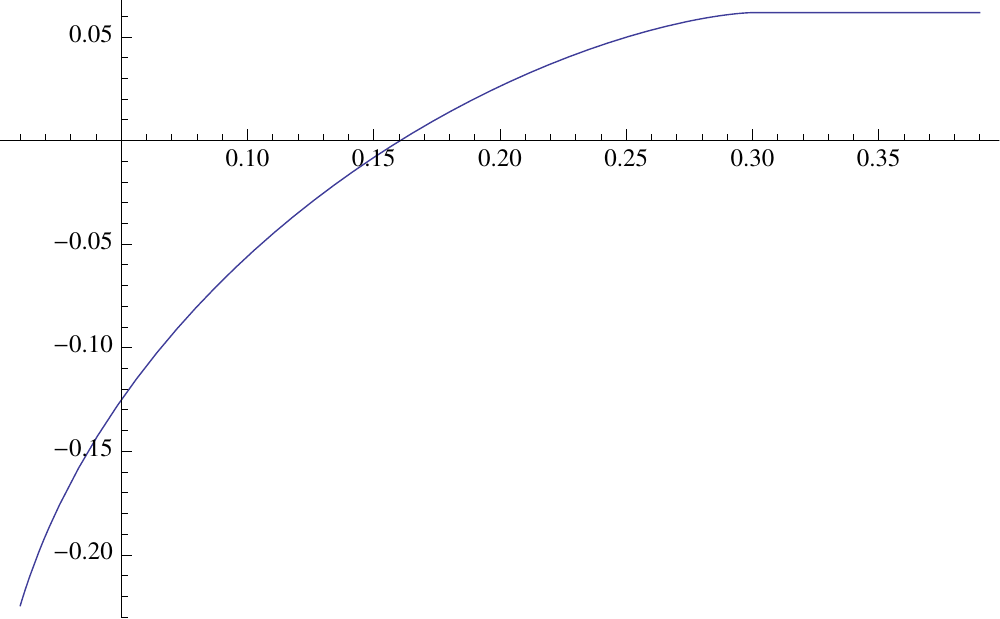}
  \caption{Graphs for Example \ref{ex:bounded_cons}.
    \emph{Top panes}: The investment strategy $\pits(w)=(1-\me^{-w})$ (left) and the
    corresponding optimal consumption $\cts(w)$ (right) for parameters:
    $r=0$, $\sharpe=0.25$, $\sigma=0.5$, $\beta=0.3$.
    \emph{Bottom panes}: The absolute risk aversion $\rho(t,c)=\rho(c)$ (left) inferred from these actions and a compatible utility function $u(0.1,c)$ (right) which is constant on $[\beta,\infty)$.
    \label{fig:ex_bounded2}
  }
\end{figure}
\end{example}

\section{Further research}\label{sec:further}

The work presented in this paper may be seen as the first step which
motivates exploration of a set of wider related questions.  Our
underpinning principle is to start with those actions which may be
observed in an investor's behaviour, and then attempt to determine
whether their actions are consistent with utility maximisation. In
this paper, we have considered two cases: a deterministic setup and a
stochastic complete market setup. In the deterministic case, our
fundamental conclusion is that observing investor's actions for any
given wealth is not enough to fully specify their utility. Risk
aversion remains unspecified. In the stochastic case, we suppose we
observe both consumption and investment. Then the assumption that the
investor is maximising utility implies that the consistency constraint
\eqref{eqn:black} holds. These two studies would have natural, and
interesting, analogues in other markets such as one period models
(where the investor can choose to consume now or in the subsequent
period), or, at the other extreme in terms of complexity,
continuous-time models which are incomplete (e.g.~a stochastic factor
model). The questions parallel to those which we have answered here
would include:
\begin{itemize}
\item is specifying an investor's consumption and investment
  strategies sufficient to determine their utility function (up to
  constants)?
\item If it is not, is the system over-specified or under-specified?
\end{itemize}
In the case where the system is over-specified, we might expect a
consistency condition such as Black's equation, \eqref{eqn:black}, and
it is interesting to ask whether there is a more general optimisation
problem such as \eqref{eq:vfdefnmod} which may correspond to the
general choice of consumption and investment.

Following on from Section \ref{sec:model_uncert}, it would be
interesting to incorporate a form of model uncertainty and ask if
agents' actions may arise from maximising \emph{some} utility under
\emph{some} dynamics of the price process.  Within our framework, the
scope for such questions was limited since we only considered
Black-Scholes dynamics for the price process. A more extensive study
could be based on the work of \citet{HeHuang:94}, see also Example 3
therein, or follow on from the research suggested above.

An even more ambitious task would be to consider an inverse problem to
the classical analysis of decision making under model uncertainty
(also called Knightian uncertainty). In that framework, one specifies
actors' preferences and a way to quantify their uncertainty about the
true price dynamics. One then asks what are the actors' optimal actions,
see e.g.\ \citet{Schied:07}, \citet{FollmerSchiedWeber:09}. The
inverse approach would start with agents' actions and try to recover
their preferences as well as their belief about model uncertainty.

Finally, a nice feature of the paper is that we have been able to go
beyond simply recovering the utility function, and have also been able
to provide characterisations of certain aspects of the agent's
behaviour (absolute and relative risk aversion) in terms of the given
data. In more complex situations, where it may not be possible to
fully recover an agent's utility function, it may still be possible to
deduce some of these related properties from the given data.

We believe the questions raised above form an exciting research programme and
we hope to pursue some of them in subsequent work.

\bibliography{Consumption} \bibliographystyle{elsart-harv}

\end{document}